\documentclass[twocolumn,prd,amssymb,aps,nofootinbib,showpacs,epsf]{revtex4}
\usepackage[usenames]{color}
\usepackage{ulem}
\usepackage{graphicx}
\usepackage{amssymb}
\usepackage{amsmath}
\usepackage{epstopdf}
\usepackage{hyperref}

\begin{document}

\title{Reconstructing the neutron-star equation of state with gravitational-wave detectors from a realistic population of inspiralling binary neutron stars}

\author{Benjamin D. Lackey$^{1, 2}$, Leslie Wade$^3$}

\affiliation{
$^1$Department of Physics, Princeton University, Princeton, NJ 08544, USA\\
$^2$Department of Physics, Syracuse University, Syracuse, NY 13244, USA\\
$^3$Department of Physics, University of Wisconsin--Milwaukee, P.O. Box 413, Milwaukee, Wisconsin 53201, USA
}

\begin{abstract}
Gravitational-wave observations of inspiralling binary neutron star systems can be used to measure the neutron-star equation of state (EOS) through the tidally induced shift in the waveform phase that depends on the tidal deformability parameter $\lambda$. Previous work has shown that $\lambda$, a function of the neutron-star EOS and mass, is measurable by Advanced LIGO for a single event when including tidal information up to the merger frequency. In this work, we describe a method for stacking measurements of $\lambda$ from multiple inspiral events to measure the EOS. We use Markov chain Monte Carlo simulations to estimate the parameters of a 4-parameter piecewise polytrope EOS that matches theoretical EOS models to a few percent. We find that, for ``realistic'' event rates ($\sim 40$ binary neutron star inspiral events per year with signal-to-noise ratio $> 8$ in a single Advanced LIGO detector), combining a year of gravitational-wave data from a three-detector network with the constraints from causality and recent high mass neutron-star measurements, the EOS above nuclear density can be measured to better than a factor of two in pressure in most cases. We also find that in the mass range $1M_\odot$--$2M_\odot$, the neutron-star radius can be measured to better than $\pm 1$~km and the tidal deformability can be measured to better than $\pm 1 \times 10^{36}$~g~cm$^2$~s$^2$ (10\%--50\% depending on the EOS and mass). The overwhelming majority of this information comes from the loudest $\sim 5$ events. Current uncertainties in the post-Newtonian waveform model, however, lead to systematic errors in the EOS measurement that are as large as the statistical errors, and more accurate waveform models are needed to minimize this error.
\end{abstract}


\maketitle

\section{Introduction}

Observations of neutron stars (NSs), consisting of nuclear matter in the ground state with densities up to several times nuclear saturation density ($\rho_{\rm nuc}\sim2.8\times 10^{14}$~g/cm$^3$), in principle provide an ideal way to measure the nuclear equation of state (EOS) that describes the pressure $p$ as a function of density $\rho$. At these densities, the most rigorous EOS constraints from NSs have come from measurements of high mass pulsars in compact binary systems, and it is now clear that the EOS must allow for NS masses $\gtrsim 2 M_\odot$~\cite{DemorestPennucciRansom2010, AntoniadisFreireWex2013}. More precise information can be obtained if the NS mass $M$ and radius $R$ are simultaneously measured, and Lindblom demonstrated explicitly that there is a one-to-one map between the relations $R(M)$ and $p(\rho)$~\cite{Lindblom1992}. 

However, various calculations of the mass and radius of NSs from available observations have produced only marginally consistent results. For example, \"Ozel~{\it et al.}\ used observations of thermonuclear bursts from x-ray binaries to measure the NS mass and radius, and found, for the three systems considered, 95\% confidence intervals that were all in the range 9--12~km at $1.4M_\odot$ (from Fig.~1 of Ref.~\cite{OzelBaymGuver2010} before stacking observations). Steiner~{\it et al.}\ also used mass and radius measurements from three x-ray bursters as well as from three quiescent low-mass x-ray binaries and found the NS radius to be 10.7--12.5~km at $1.4M_\odot$ (from the range of 95\% confidence intervals in Tables~7 and~8 of Ref~\cite{SteinerLattimerBrown2010}). Finally, Guillot~{\it et al.}\ measured the mass and radius for five quiescent low-mass x-ray binaries, and found that the NS radius was 7.6--10.4km (90\% confidence) assuming the radius was an approximately constant function of mass~\cite{GuillotServillatWebb2013}.

Gravitational-wave (GW) observations of NSs in inspiralling compact binaries, on the other hand, are sensitive only to the density profile of NSs, and are therefore free of the model-dependent uncertainties in the emission and absorption mechanisms that plague electromagnetic observations. In the next few years, observing runs will begin for second generation detectors, including the two Advanced LIGO (aLIGO) detectors~\cite{Harry2010} and Advanced Virgo (aVirgo)~\cite{Acernese2009}, and design sensitivity will likely be reached by the end of the decade~\cite{AasiAbadieAbbott2013}. In addition, \mbox{KAGRA}~\cite{Somiya2012} (formerly LCGT) and possibly LIGO-India~\cite{IyerSouradeepUnnikrishnan2011} will come online a few years later. 

The EOS information provided by these detectors comes mainly from tidal interactions during the inspiral of binary neutron star (BNS) and black hole-neutron star (BHNS) systems that induce quadrupolar deformations in the NSs~\cite{FlanaganHinderer2008}. In the quasistationary approximation, the quadrupole moment $Q_{ij}$ of one star depends on the tidal field $\mathcal{E}_{ij}$ from the monopole of the other star through the relation $Q_{ij} = -\lambda \mathcal{E}_{ij}$. Here, $\lambda$ is the EOS dependent tidal deformability and is related to the NS's dimensionless Love number $k_2$, first calculated in Ref.~\cite{Hinderer2008}, and radius $R$ through the relation $\lambda = \frac{2}{3G} k_2 R^5$, where $G$ is the gravitational constant.

Because the tidal effect is a strong function of the mass ratio, and observed BNS events are likely to be significantly closer, this tidal interaction is likely to be seen for only BNS systems~\cite{PannaraleRezzollaOhmeRead2011}. Additional information may also come from modes in the post-merger remnant of BNS systems~\cite{StergioulasBausweinZagkouris2011, ClarkBausweinCadonati2014} as well as the damping of quasi-normal modes for low mass, high spin BHNS systems~\cite{LackeyKyutokuShibata2014}.

The measurability of tidal parameters in BNS systems with aLIGO for GW frequencies below 450Hz (prior to the last $\sim 20$ GW cycles before merger where higher order corrections to the tidal effect become important) was first examined for polytropic EOSs in Ref.~\cite{FlanaganHinderer2008} and for theoretical hadronic and quark matter EOSs in Ref.~\cite{HindererLackeyLangRead2010}.  These studies found that for a single aLIGO detector, tidal interactions were only observable during this early inspiral stage for stiff EOSs, NS masses below 1.4~$M_\odot$, and for rare, nearby sources with signal-to-noise ratios (SNR) $\gtrsim 30$. Damour {\it et al.}, however, found that when also including EOS information from the last $\sim 20$ GW cycles prior to contact, tidal parameters are in fact observable even with more common SNRs of $\sim 16$~\cite{DamourNagarVillain2012}. 

These studies relied on the Fisher matrix approximation for parameter estimation which assumes the distribution of parameters follows a multivariate Gaussian. Recent works using a fully Bayesian analysis for the three-detector aLIGO-aVirgo network have confirmed that the tidal signal is indeed measurable when including the entire inspiral up to merger~\cite{DelPozzoLiAgathos2013, WadeCreightonOchsner2014}. In particular Del Pozzo {\it et al.}\ used a method to stack tens of observations of a large number of BNS inspiral events to measure $\lambda(M)$~\cite{DelPozzoLiAgathos2013}. By parametrizing $\lambda(M)$ with a linear fit, they found that $\lambda$ could be measured to $\pm 10\%$ at $1.4M_\odot$, but the mass dependence of $\lambda$ could not be found. This was true even though they used an unrealistically large NS mass range of $1M_\odot$--$2M_\odot$ for their simulated population.

In this paper we will demonstrate the advantages of parameterizing the EOS instead of $\lambda(M)$. The main benefit comes from including prior knowledge of the EOS that would be more difficult to incorporate into the $\lambda(M)$ fit. As will be discussed in Section~\ref{sec:eos}, the requirements that the EOS is monotonic and causal provide the simple constraint that the slope of the EOS is in the range $0 \le dp/d\epsilon \le c^2$, where $\epsilon$ is the energy density and $c$ is the speed of light. Additionally, the constraints from the observed $2M_\odot$ NSs can be included by simply rejecting any EOS parameters that lead to a maximum NS mass below $2M_\odot$. We will find that incorporating this information allows us to make significantly stronger statements about the EOS, radius, and tidal deformability, and this is true for a narrower, more realistic range of NS masses than simulated by Del Pozzo {\it et al.}~\cite{DelPozzoLiAgathos2013}.

The BNS inspiral signal is almost exactly characterized by the mass, spin, and tidal deformability of each NS, and is therefore free from the intrinsic variability and uncertainties that bias electromagnetic measurements of the mass and radius. Currently unknown terms in the post-Newtonian (PN) waveform model, however, will lead to systematic errors in recovering the tidal deformabilities. Using the Fisher matrix approximation~\cite{Favata2014, YagiYunes2014} as well as a Markov chain Monte Carlo (MCMC) Bayesian analysis~\cite{WadeCreightonOchsner2014}, it was found that the systematic error in $\lambda$ from waveform uncertainties can be as large as the statistical error for a single BNS inspiral observation. When stacking observations the problem becomes worse, as the statistical errors decrease with more observations but the systematic errors do not, and we will discuss how systematic errors impact the recovery of the EOS.

We organize the paper as follows. We describe the BNS waveform model in Section~\ref{sec:waveform} and the EOS parameterization in Section~\ref{sec:eos}. Our Bayesian method for estimating EOS parameters is derived in Section~\ref{sec:bayes}. We then show our results for a range of observation scenarios in Section~\ref{sec:results} and the impact from waveform uncertainties in Section~\ref{sec:systematic}. Finally, we summarize our results in Section~\ref{sec:discussion}.

\section{The BNS inspiral signal}
\label{sec:waveform}

The strain $h$ observed in a detector from a GW is related to the two polarizations of the GW, $h_+$ and $h_\times$, by the detector's antenna beam pattern response, $F_+$ and $F_\times$, through the relation
\begin{equation}
h = F_+ h_+ + F_\times h_\times.
\end{equation}
$F_+$ and $F_\times$ depend on the sky position (given by the right ascension $\alpha$ and declination $\delta$) and GW polarization angle $\psi$ of the source.

For a BNS system in a quasicircular inspiral with an inclination angle $\iota$, comoving (transverse) distance $d$, and component masses $m_1$ and $m_2$, the two polarizations of the GW are
\begin{eqnarray}
\label{eq:hoftp}
h_+(t)&=&-\frac{4G\eta M}{c^2 d}\left(\frac{1+\cos^2\iota}{2}\right)x(t)\cos\left[2\Phi(t)\right],\\
\label{eq:hoftc}
h_\times(t)&=&-\frac{4G\eta M}{c^2 d}\cos\iota\mbox{ }x(t)\sin\left[2\Phi(t)\right],
\end{eqnarray}
where we have restricted the amplitude to the leading order and only keep the leading harmonic. Here, $M=m_1+m_2$ is the total mass, $\eta=m_1 m_2/M^2$ is the symmetric mass ratio, $\Phi$ is the orbital phase, and $x$ is the standard PN order parameter defined by $x = \left( \frac{GM}{c^3}\frac{d\Phi}{dt} \right)^{2/3} = \left( \frac{\pi GM f}{c^3} \right)^{2/3}$, where $f$ is the source frame GW frequency of the leading harmonic.

The time evolution of $\Phi(t)$ and $x(t)$ are evaluated from the PN expressions for the energy $E$ and luminosity ${\cal L}$ via the energy balance requirement $dE/dt=-{\cal L}$:
\begin{eqnarray}
\label{eq:dPhidt}
\frac{d\Phi(t)}{dt}&=&\frac{c^3 x^{3/2}}{G M},\\
\label{eq:dxdt}
\frac{d x(t)}{dt}&=&\frac{-\mathcal{L}}{dE/dx}.
\end{eqnarray}
The energy and luminosity have the schematic form
\begin{align}
E &= - \frac{1}{2} c^2 M \eta x \left[1 + e_\text{PP-PN}(x; \eta) + e_\text{Tidal}(x; \eta, \Lambda_1, \Lambda_2)\right], \\
\mathcal{L} &= \frac{32}{5} \frac{c^5}{G} \eta^2 x^5 \left[ 1 + l_\text{PP-PN}(x; \eta) + l_\text{Tidal}(x; \eta, \Lambda_1, \Lambda_2)\right],
\end{align}
where we write the tidal terms as functions of the dimensionless quantities $\Lambda_i = G\lambda_i \left(\frac{c^2}{G m_i}\right)^5$.

The point-particle PN terms for nonspinning systems depend only on $\eta$ and $x$. The energy $e_\text{PP-PN}$ was recently calculated to 4PN order~\cite{BiniDamour2013, FoffaSturani2014}, and the luminosity $l_\text{PP-PN}$ is known to 3.5PN order~\cite{Blanchet2014Review}. For consistency with previous works~\cite{WadeCreightonOchsner2014, DelPozzoLiAgathos2013, Favata2014, YagiYunes2014}, we keep the point-particle PN corrections to 3.5PN order.

The leading tidal terms begin at the same order as 5PN point-particle terms, and although the tidal potential is known to the next-to-next-to-leading-order term~\cite{BiniDamourFaye2012}, we only include the leading and next-to-leading order terms for consistency with previous works~\cite{WadeCreightonOchsner2014, DelPozzoLiAgathos2013, Favata2014, YagiYunes2014}. These terms are~\cite{VinesFlanaganHinderer2011}
\begin{widetext}
\begin{align}
\begin{split}
e_\text{Tidal} &= -\frac{9\eta}{2} \left[ (1-3\eta)(\Lambda_1+\Lambda_2) + \sqrt{1-4\eta}(1-\eta)(\Lambda_1-\Lambda_2) \right]x^5 \\
& - 22\eta \left[ \left(1-4\eta+\frac{19\eta^2}{8}\right)(\Lambda_1+\Lambda_2) + \sqrt{1-4\eta}\left(1-2\eta+\frac{3\eta^2}{8}\right)(\Lambda_1-\Lambda_2) \right]x^6,
\end{split} \\
\begin{split}
l_\text{Tidal} &= 3 \left[ (1-2\eta-4\eta^2)(\Lambda_1+\Lambda_2) + \sqrt{1-4\eta}(1-2\eta^2)(\Lambda_1-\Lambda_2) \right]x^5 \\
& - \frac{22}{7} \left[ \left(1-\frac{15\eta}{176}+\frac{147\eta^2}{88}-\frac{1547\eta^3}{44}\right)(\Lambda_1+\Lambda_2) + \sqrt{1-4\eta}\left(1+\frac{337\eta}{176}+\frac{15\eta^2}{2}-\frac{1085\eta^3}{88}\right)(\Lambda_1-\Lambda_2) \right]x^6,
\end{split}
\end{align}
\end{widetext}
where $m_1 \ge m_2$.

As described in detail in Ref.~\cite{WadeCreightonOchsner2014}, there are several methods, first cataloged in Ref.~\cite{DamourIyerSathyaprakash2001}, to determine the phase evolution of the binary from the above expressions for its energy and luminosity, and we will use three of them.  In the TaylorT1 method, one numerically integrates Eqs.~\eqref{eq:dPhidt} and~\eqref{eq:dxdt} to find $\Phi(t)$ and $x(t)$ up to the time and phase constants $t_c$ and $\phi_c$. The TaylorT4 method is a slight variation where one re-expands the ratio $-{\cal L}/E^\prime$ in Eq.~\eqref{eq:dxdt} then truncates the series at the highest known PN order (here 3.5PN) before numerically integrating. Finally in the TaylorF2 method, one approximates the Fourier transform of the waveform
\begin{equation}
\tilde{h}(f)=\int_{-\infty}^{+\infty} h(t) e^{-2\pi i f t}dt
\end{equation}
with the stationary phase approximation. The result, to leading order in the amplitude, is
\begin{align}
\label{eq:hoffp}
\tilde{h}_+(f) &= \sqrt{\frac{5}{24}}\frac{G^2}{c^5}\frac{{\cal M}^{5/6}(Gf/c^3)^{-7/6}}{\pi^{2/3}d}\left(\frac{1+\cos^2\iota}{2}\right)e^{-i\Psi(f)},\\
\label{eq:hoffc}
\tilde{h}_\times(f) &= -i\sqrt{\frac{5}{24}}\frac{G^2}{c^5}\frac{{\cal M}^{5/6}(Gf/c^3)^{-7/6}}{\pi^{2/3}d}\cos\iota\mbox{ }e^{-i\Psi(f)}.
\end{align}
Here, $\mathcal{M} = (m_1 m_2)^{3/5}/M^{1/5}$ is the chirp mass, and, in the TaylorF2 approach, the phase has the analytic form 
\begin{equation}
\begin{split}
\Psi(f) &= 2\pi f t_c - 2\phi_c -\frac{\pi}{4}\\
& +\frac{3}{128\eta x^{5/2}}\left[1+\psi_\text{PP-PN}(x; \eta)+\psi_\text{Tidal}(x; \eta, \Lambda_1, \Lambda_2)\right],
\end{split}
\end{equation}
where the point-particle terms $\psi_{\rm PP-PN}$ are provided in Ref.~\cite{BuonannoIyerOchsner2009}. As was demonstrated in Refs.~\cite{FlanaganHinderer2008, Favata2014, WadeCreightonOchsner2014}, the individual tidal parameters $\Lambda_1$ and $\Lambda_2$ are highly correlated, so it is instead easier to reparameterize the tidal contribution to the phase $\psi_\text{Tidal}$ in terms of the linear combinations
\begin{align}
\begin{split}
\label{eq:LT}
\tilde{\Lambda}&=\frac{8}{13}\left[\left(1+7\eta-31\eta^2\right)\left(\Lambda_1+\Lambda_2\right)\right.\\
&+ \left.\sqrt{1-4\eta}\left(1+9\eta-11\eta^2\right)\left(\Lambda_1-\Lambda_2\right)\right],
\end{split}\\
\begin{split}
\label{eq:dLT}
\delta\tilde{\Lambda}&=\frac{1}{2}\left[\sqrt{1-4\eta}\left(1-\frac{13272}{1319}\eta+\frac{8944}{1319}\eta^2\right)\left(\Lambda_1+\Lambda_2\right) \right .\\
& + \left . \left(1-\frac{15910}{1319}\eta+\frac{32850}{1319}\eta^2+\frac{3380}{1319}\eta^3\right)\left(\Lambda_1-\Lambda_2\right)\right].
\end{split}
\end{align}
The tidal contribution then takes the simple form
\begin{equation}
\label{eq:psi_tidal}
\psi_\text{Tidal} = -\frac{39}{2}\tilde{\Lambda} x^5+ \left(-\frac{3115}{64}\tilde{\Lambda}+\frac{6595}{364}\sqrt{1-4\eta}\mbox{ }\delta\tilde{\Lambda}\right) x^6.
\end{equation}
The two terms containing $\tilde\Lambda$ are significantly larger than the term containing $\delta\tilde\Lambda$, and, as was previously found, $\delta\tilde\Lambda$ is not measurable with aLIGO~\cite{WadeCreightonOchsner2014}. This waveform model can therefore be expressed in terms of the 11 parameters $\vec\theta = \{d, \alpha, \delta, \psi, \iota, t_c, \phi_c, m_1, m_2, \tilde\Lambda, \delta\tilde\Lambda\}$.

For all versions of the PN waveform, we cut off the inspiral at the GW frequency corresponding to the Schwarzschild innermost stable circular orbit (ISCO) $f_{\rm ISCO} = c^3/(6^{3/2}\pi G M)$.  However, for large NS radii corresponding to a stiff EOS, NSs can merge before reaching $f_{\rm ISCO}$. For simplicity, we also choose to examine nonspinning BNS systems, although it has been shown that not including spin parameters can noticeably bias parameter estimation even for systems with relatively small spin magnitudes \cite{Favata2014}. A parallel investigation which studies how the choice of high-frequency cutoffs and spin can affect the estimation of tidal parameters is nearing completion \cite{Agathos2014}.

\section{The EOS}
\label{sec:eos}

Because the EOS can be calculated from the $R(M)$~\cite{Lindblom1992} or $\lambda(M)$ relations~\cite{LindblomIndik2012, LindblomIndik2014}, we can choose to parameterize either $R(M)$, $\lambda(M)$ or $p(\rho)$ if we want to reconstruct the EOS. As Del Pozzo {\it et al.}\ found by parameterizing $\lambda(M)$ with a linear fit, it is difficult to accurately measure $\lambda(M)$ over a wide range of masses with BNS inspiral observations~\cite{DelPozzoLiAgathos2013}. Instead, they found that $\lambda$ could only be accurately measured for a specific fiducial mass which they chose to be $1.4M_\odot$. This is because, when parameterizing the function $\lambda(M)$, there is little {\it a priori} information about the allowed functional form of $\lambda(M)$. Lattimer and Steiner found similar results for mass and radius observations (Section~4.1 of Ref.~\cite{LattimerSteiner2014}) when they parameterized $R(M)$ instead of the EOS. In contrast, it is much simpler to place useful constraints on the functional form of the EOS fit, and this allows one to make significantly stronger statements about the EOS and the behavior of the $\lambda(M)$ and $R(M)$ curves.

\subsection{Current EOS constraints}
\label{sec:eosconstraints}

The {\it a priori} constraints on the EOS that we will use are:
\begin{enumerate}
\item The EOS can be considered known below a certain density $\rho_0$, and we use a fixed EOS below that density.
\item The EOS must be a monotonically increasing function ($dp/d\epsilon \ge 0$, where $\epsilon$ is the energy density) to satisfy thermodynamic stability.
\item The speed of sound $v_s = \sqrt{dp/d\epsilon}$ must be less than the speed of light, ensuring causality. 
\item The EOS must allow for a maximum NS mass greater than observed masses. Recent pulsar mass measurements in two neutron star-white dwarf binaries have provided convincing evidence that the maximum NS mass is $\gtrsim 2M_\odot$. The pulsar J1614-2230 was found to have a mass of $1.97\pm0.04M_\odot$ ($1\sigma$ confidence)~\cite{DemorestPennucciRansom2010}, and the pulsar J0348+0432 was found to have a mass of $2.01\pm0.04M_\odot$ ($1\sigma$ confidence)~\cite{AntoniadisFreireWex2013}. We will take the $2\sigma$ lower bound of $1.93M_\odot$ on the mass of J0348+0432 as a solid lower bound on the maximum NS mass.
\end{enumerate}
Although other constraints exist, such as the constraint that the maximum mass-shedding (Kepler) frequency be greater than observed spin frequencies, these turn out to be less useful, so we will focus on the ones listed above. See Ref.~\cite{ReadLackey2009} for a discussion of other constraints.

\subsection{Choice of parameterization}
\label{sec:eosparam}

We seek a parameterized EOS that will satisfy the above constraints and also have enough freedom to accurately fit the true EOS. Several choices have been presented in the literature. Read {\it et al.}\ examined various types of piecewise polytropes, defined below, and found that a four-parameter fit could adequately match a wide range of theoretical models~\cite{ReadLackey2009}. Steiner {\it et al.}\ used a model with four nuclear parameters around nuclear saturation density and an additional 4 parameters at higher densities to define a 2-piece polytrope with variable dividing densities~\cite{SteinerLattimerBrown2010}, and they also examined several variations~\cite{SteinerLattimerBrown2013}. Finally, Lindblom constructed a spectral expansion of $\gamma(\epsilon) = d\log p / d\log \epsilon$ that appears to converge to tabulated EOS models with fewer parameters than the 4-parameter piecewise polytrope constructed by Read {\it et al.}~\cite{Lindblom2010}. All of these models can be made to satisfy the constraints in Section~\ref{sec:eosconstraints}. However, because the 4-parameter piecewise polytrope~\cite{ReadLackey2009} has been more commonly used in the GW literature, we will focus on it in this paper, and leave a detailed comparison of how the results depend on the choice of parameterization to future work.

For each piece of a piecewise polytrope, the pressure $p$ in the rest-mass density interval $\rho_{i-1} < \rho < \rho_i$ is defined by
\begin{equation}
p(\rho) = K_i \rho^{\Gamma_i},
\end{equation}
where $\Gamma_i$ is the adiabatic index and the constant $K_i$ is chosen such that $p$ is continuous at the boundary $\rho_{i-1}$. For the core of the star, we use 3 polytropes with adiabatic indices $\Gamma_1$, $\Gamma_2$, and $\Gamma_3$ separated by the fixed dividing densities $\rho_1 = 10^{14.7}$~g/cm$^3$ and $\rho_2 = 10^{15}$~g/cm$^3$. These fixed dividing densities were chosen to minimize the least-squares error between the piecewise polytrope fit and a set of 34 tabulated theoretical EOSs~\cite{ReadLackey2009}. In addition to the three adiabatic indices, we also require an additional parameter to determine the overall pressure scaling which we choose to be $p_1 = p(\rho_1)$, the pressure at the first dividing density. We will therefore use as our four EOS parameters $\vec E = \{\log(p_1), \Gamma_1, \Gamma_2, \Gamma_3\}$.

For the lower density crust of the NS we use a 4-piece polytrope fit to the SLy EOS given by Table~II of Ref.~\cite{ReadLackey2009}\footnote{The values for $K_i$ in Table~II of Ref.~\cite{ReadLackey2009} incorrectly give the pressure in mass-density units $p/c^2$. The values of $K_i$ in the table should therefore be multiplied by $c^2$.}, and we note that the particular choice of the crust EOS effects the results at less than a percent level. We join the fixed crust EOS and parameterized core EOS together at the density $\rho_0$ where the two EOSs intersect. This usually occurs below $\rho_{\rm nuc}$, and we reject any set of EOS parameters where the joining density is below the start of the last polytrope piece for the crust at $2.63\times 10^{12}$~g/cm$^3$, where we consider the EOS to be known. Finally, given this EOS, the energy density $\epsilon$, used for solving the stellar structure equations that determine the radius and tidal deformability, can be evaluated by integrating the first law of thermodynamics
\begin{equation}
d\frac{\epsilon}{\rho} = - p d\frac{1}{\rho}.
\end{equation}

We note that for NSs around $1.4M_\odot$ found in BNS systems, the central density is close to or just above the last dividing density $\rho_2 = 10^{15}$~g/cm$^3$ for most EOS parameters~\cite{ReadLackey2009}. Information about $\Gamma_3$ therefore cannot come from BNS inspiral observations alone. However, when used in conjunction with the priors from causality and high mass NS observations, the inspiral will help constrain $\Gamma_3$ and the EOS above $\rho_2$.

\begin{figure}[!htb]
\begin{center}
\includegraphics[width=3.2in]{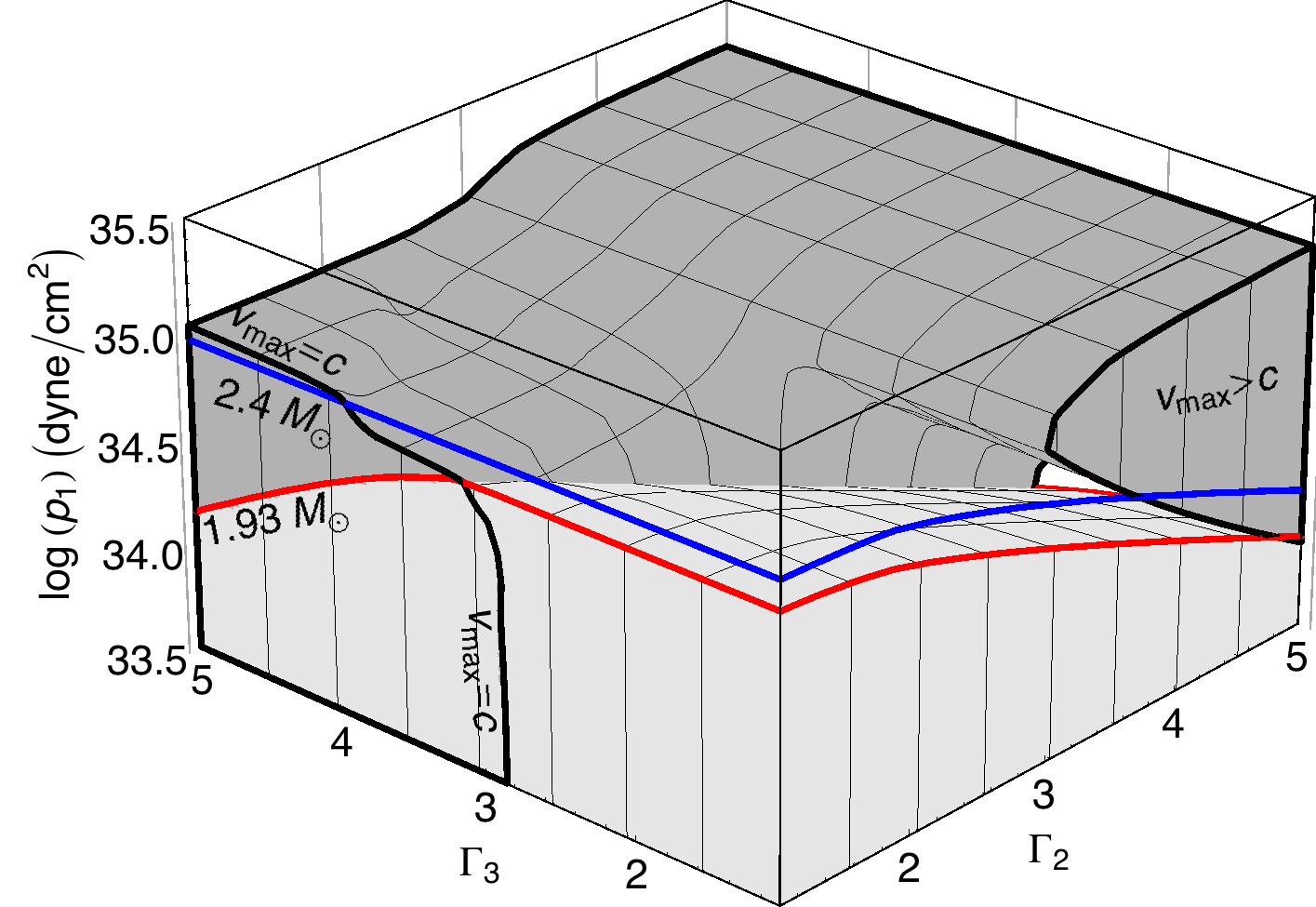}
\caption{The joint constraint imposed by causality and the existence of a $1.93M_\odot$ NS. In the dark shaded volume outlined in black, the EOS parameters allow a NS to have $v_s>c$ at some density below $\rho_{\rm c, max}$, and are therefore ruled out. In the light shaded volume outlined in red, the EOS parameters are ruled out because they don't allow a NS to have a maximum mass of at least $1.93M_\odot$. Also outlined in blue is the surface where the maximum mass is $2.4M_\odot$. A hypothetical $2.4M_\odot$ NS observation would rule out EOS parameters below this surface.}
\label{fig:mvs3d}
\end{center}
\end{figure}

In Fig.~\ref{fig:mvs3d}, we show the constraints placed on the parameter space by the causality requirement and the existence of a NS with a mass of at least $1.93M_\odot$. Although the causality requirement must hold for all densities, we have only assumed that the piecewise polytrope used here is a sufficiently good fit to the true EOS for densities below $\rho_{\rm c, max}$, the central density of the maximum mass NS. Above this density the EOS could take some other form that is not well described by the expression $p = K_3 \rho^{\Gamma_3}$. We have therefore used the weaker constraint of excluding EOS parameters that result in $v_s > c$ below $\rho_{\rm c, max}$ but accepting those parameters if $v_s > c$ above $\rho_{\rm c, max}$ as discussed in Ref.~\cite{ReadLackey2009}. Because most of the mass in a NS at its maximum allowed mass is above $\rho_1$, the maximum mass and maximum speed of sound are mostly independent of $\Gamma_1$, the adiabatic index below $\rho_1$. We therefore show the constraint in the 3-parameter $\{\log(p_1), \Gamma_2, \Gamma_3\}$ subspace with $\Gamma_1 = 2.1$, the value that restricts the other parameters the least. 

In our analysis below, we will also use the constraints $\log(p_1/(\text{dyne cm}^{-2})) \in [33.5, 34.5]$, $\Gamma_1 \in [1.4, 5]$, $\Gamma_2 \in [1, 5]$, and $\Gamma_3 \in [1, 5]$. We also use the above requirement that $\rho_0 \ge 2.63\times 10^{12}$~g/cm$^3$ which restricts small values of $\Gamma_1$ when $\log(p_1)$ is large as shown in Fig.~4 of Ref.~\cite{ReadLackey2009}. These boundaries are large enough to incorporate the 34 EOSs considered in Ref.~\cite{ReadLackey2009}, and are also large enough that they have a minimal impact on the results below.

\subsection{Comparison between parameterized and theoretical EOSs}

\begin{figure}[!htb]
\begin{center}
\includegraphics[width=3.2in]{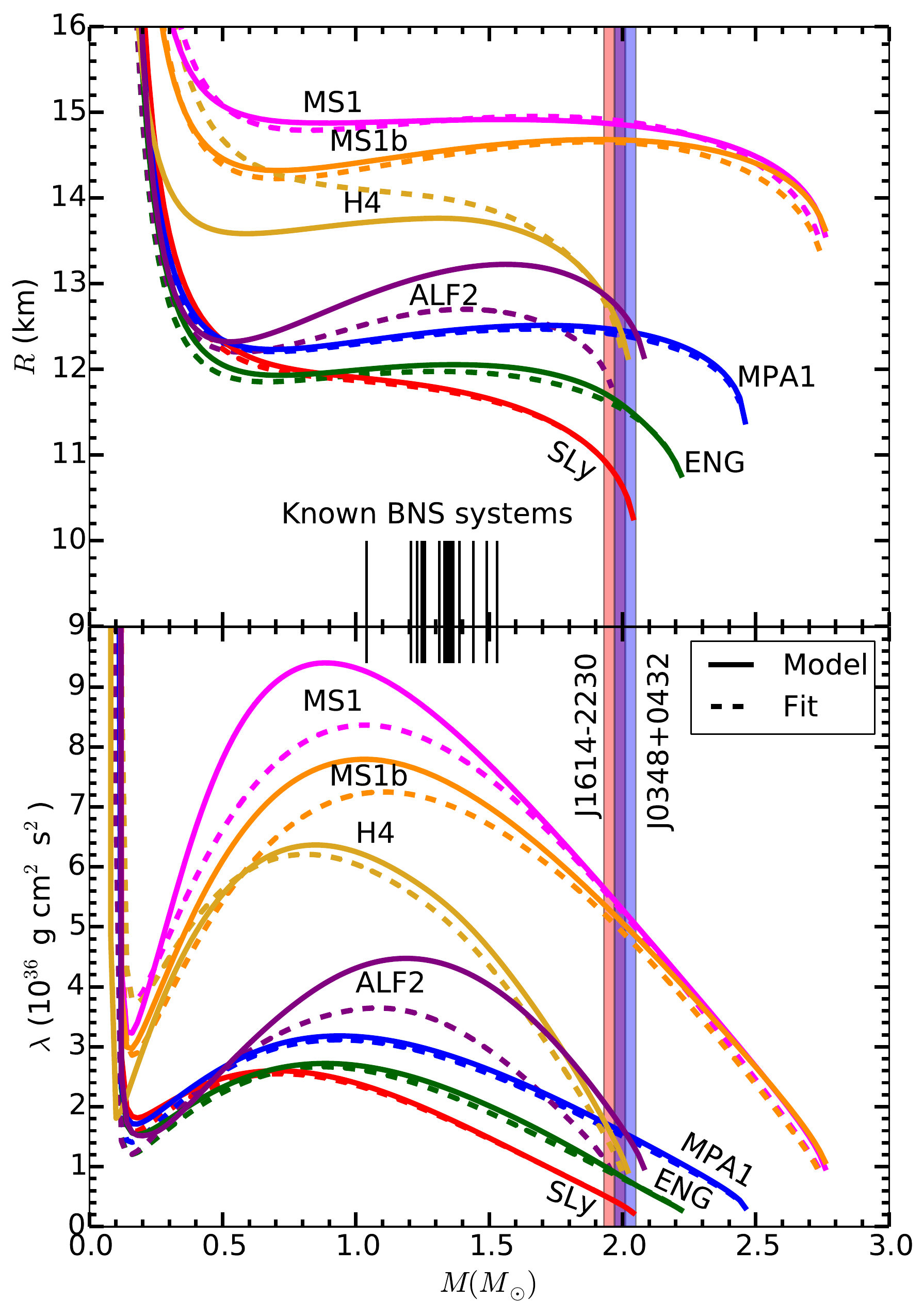}
\caption{Radius and tidal deformability of tabulated EOS models (solid) and the least-squares piecewise-polytrope fits (dashed) to those tabulated models given in Table~\ref{tab:eosfit}. The 20 vertical lines represent the most likely NS masses of the 10 known BNS systems~\cite{Lattimer2012}. Some of these masses, however, have significant uncertainties. The overlapping vertical bands represent the $1\sigma$ uncertainty in the masses of the pulsars J1614-2230 ($1.97\pm0.04M_\odot$)~\cite{DemorestPennucciRansom2010} and J0348+0432 ($2.01\pm0.04M_\odot$)~\cite{AntoniadisFreireWex2013}, both in neutron star-white dwarf binaries.}
\label{fig:rlofm}
\end{center}
\end{figure}

The piecewise-polytrope fit was originally constructed to match a wide range of theoretical EOSs whether or not they satisfied the then current constraints~\cite{ReadLackey2009}. In this paper we will examine the 7 EOSs that satisfy the causality constraint for densities up to $\rho_{\rm c, max}$ and also have maximum masses above $2M_\odot$. We list these EOSs in Table~\ref{tab:eosfit} along with the piecewise-polytrope parameters that minimize the least-squares residual~\cite{ReadLackey2009}. We also list several NS properties and the associated errors in reproducing them with the fit. In Fig.~\ref{fig:rlofm}, we compare the radius and tidal deformability for these EOSs with their least-squares fits.  The error in $R$ and $\lambda$ above $1M_\odot$ for the EOS fits is usually less than 10\%, except for the ALF2 EOS.

\begin{table*}[!htb]
\begin{center}
\caption{
Comparison between tabulated EOS models and their best fits. The parameters $\log(p_1/\text{(dyne~cm$^{-2}$)})$, $\Gamma_1$, $\Gamma_2$, and $\Gamma_3$ are the values that minimize the least-squares residual defined in Ref.~\cite{ReadLackey2009}. Observables for the tabulated EOSs are also shown; $v_{\rm s,max}$ ($c$) is the maximum speed of sound below $\rho_{\rm c, max}$, $M_{\rm max}$ ($M_\odot$) is the maximum mass, $R_{1.4}$ (km) is the radius of a $1.4~M_\odot$ NS, and $\lambda_{1.4}$ ($10^{36}$~g~cm$^2$~s$^2$) is the tidal deformability of a $1.4~M_\odot$ NS. The percent error for each observable when using the tabulated EOS $(O_{\rm tab})$ versus the best-fit parameterized EOS $(O_{\rm fit})$ is also given by $(O_{\rm fit}/O_{\rm tab}-1)100$.
}
\begin{tabular}{l|ccccl|rr|rr|rr|rr}
\hline\hline
EOS & $\log(p_1)$ & $\Gamma_1$ & $\Gamma_2$ & $\Gamma_3$ & residual &
$v_{s, \rm max}$ & \% & $M_{\rm max}$ & \% & $R_{1.4}$ & \% & $\lambda_{1.4}$ & \% \\
\hline
SLy&34.384&3.005&2.988&2.851&0.0020& 0.989&1.41& 2.049&0.02& 11.736&-0.21&1.69&-1.10\\
ENG&34.437&3.514&3.130&3.168&0.015& 1.000&10.71& 2.240&-0.05& 12.059&-0.69&2.20&-4.93\\
MPA1&34.495&3.446&3.572&2.887&0.0081& 0.994&4.91& 2.461&-0.16& 12.473&-0.26&2.78&-2.47\\
MS1&34.858&3.224&3.033&1.325&0.019& 0.888&12.44& 2.767&-0.54& 14.918&0.06&8.13&-4.17\\
MS1b&34.855&3.456&3.011&1.425&0.015& 0.889&11.38& 2.776&-1.03& 14.583&-0.32&7.28&-4.69\\
H4&34.669&2.909&2.246&2.144&0.0028& 0.685&4.52& 2.032&-0.85& 13.774&1.34&5.12&-5.40\\
ALF2&34.616&4.070&2.411&1.890&0.043& 0.642&1.50& 2.086&-5.26& 13.188&-3.66&4.27&-24.34\\
\hline\hline
\end{tabular}
\label{tab:eosfit}
\end{center}
\end{table*}

\section{Bayesian inference of EOS parameters}
\label{sec:bayes}

\subsection{Derivation}

Given the GW model and EOS fit described in Sections~\ref{sec:waveform} and~\ref{sec:eos}, we will describe here a method for estimating the EOS parameters from a set of GW observations. We want to find the posterior density function (PDF) $p(\vec E,\vec\theta_1,\dots,\vec\theta_n | D,\mathcal{H},\mathcal{I})$ for the universal EOS parameters $\vec E$ that are common to all NSs~\cite{Glendenning1996book} and the waveform parameters $\vec\theta_i$ that are unique to each of the $n$ BNS inspiral events\footnote{The EOS parameters are related to the waveform parameters through the relations $\tilde\Lambda_i = \tilde\Lambda(m_{1i}, m_{2i}, \vec E)$ and $\delta\tilde\Lambda_i = \delta\tilde\Lambda(m_{1i}, m_{2i}, \vec E)$, and it would be possible to use the EOS parameters instead of $\tilde\Lambda_i$ and $\delta\tilde\Lambda_i$ as part of the waveform parameters. However, it is simpler to not modify existing parameter estimation codes, and treat them as separate parameters until Eq.~\eqref{eq:lambdaeosrelation}.}. The data $D = \vec d_1,\dots, \vec d_n$ is composed of the data streams $\vec d_i(t)$ from the GW detector network for each of the $n$ inspiral events. $\mathcal{H}$ represents the waveform model, chosen here to be the TaylorF2 waveform with parameters $\vec\theta = \{d, \alpha, \delta, \psi, \iota, t_c, \phi_c, m_1, m_2, \tilde\Lambda, \delta\tilde\Lambda\}$, as well as the EOS model, chosen to be the four-parameter piecewise polytrope with parameters $\vec E = \{\log(p_1), \Gamma_1, \Gamma_2, \Gamma_3\}$. $\mathcal{I}$ represents the background information for the waveform and EOS parameters. The PDF is given by Bayes' theorem 
\begin{equation}
\label{eq:bayes}
\begin{split}
p(\vec E,\vec\theta_1,\dots,\vec\theta_n | D,\mathcal{H},\mathcal{I})
&= p(\vec E,\vec\theta_1,\dots,\vec\theta_n | \mathcal{H},\mathcal{I}) \\
& \times\frac{ p(D | \vec E,\vec\theta_1,\dots,\vec\theta_n,\mathcal{H},\mathcal{I}) }{ p(D | \mathcal{H},\mathcal{I}) }.
\end{split}
\end{equation}
The quantity $p(\vec E,\vec\theta_1,\dots,\vec\theta_n | \mathcal{H},\mathcal{I})$ is the prior probability density for the EOS and waveform parameters, $p(D | \vec E,\vec\theta_1,\dots,\vec\theta_n,\mathcal{H},\mathcal{I})$ is the likelihood, and the normalization constant $p(D | \mathcal{H},\mathcal{I})$, which we will not need to calculate here, is the evidence. We will then want to integrate over the waveform parameters to obtain a marginalized PDF for just the EOS parameters
\begin{equation}
\label{eq:margEOS}
p(\vec E | D,\mathcal{H},\mathcal{I}) = \int d\vec\theta_1 \dots d\vec\theta_n p(\vec E,\vec\theta_1,\dots,\vec\theta_n | D,\mathcal{H},\mathcal{I}).
\end{equation}
  
This $11n$-dimensional integral is not easily computed, so we will decompose it into blocks and calculate it in a two step procedure. In the first step, we use an existing MCMC algorithm to sample the posterior for the parameters $\vec\theta_i$ of each BNS event, then marginalize over the extrinsic/nuisance parameters $\vec\theta_{\rm ex} = \{d, \alpha, \delta, \psi, \iota, t_c, \phi_c, \delta\tilde\Lambda\}$\footnote{$\delta\tilde\Lambda$ is an intrinsic parameter that in principle provides additional EOS information. However, it is unmeasurable with aLIGO~\cite{WadeCreightonOchsner2014}, so we treat it as a nuisance parameter and group it with the extrinsic parameters when we marginalize over them.} to obtain a quasilikelihood (Eq.~\eqref{eq:quasilike}) for the intrinsic parameters $\vec\theta_{\rm in} = \{m_1, m_2, \tilde\Lambda\}$ that are relevant for measuring the EOS. In the second step, we evaluate Eq.~\eqref{eq:margEOS} by constructing a joint likelihood for the $n$ BNS events from the quasilikelihoods for each event, then re-express $\tilde\Lambda_i$ in terms of the EOS parameters and masses, and marginalize over the masses in Eq.~\eqref{eq:margEOS2} using another MCMC algorithm. We obtain these expressions from Bayes' theorem (Eq.~\eqref{eq:bayes}) as follows.  

The prior can be decomposed into EOS-parameter and waveform-parameter parts using the product rule $p(\vec x, \vec y) = p(\vec x) p(\vec y | \vec x) $ as well as the fact that the $n$ sets of waveform parameters $\vec\theta_i$ are independent
\begin{equation}
p(\vec E,\vec\theta_1,\dots,\vec\theta_n | \mathcal{H},\mathcal{I}) = p(\vec E | \mathcal{H},\mathcal{I}) \prod_{i=1}^n p(\vec\theta_i | \vec E, \mathcal{H},\mathcal{I}).
\end{equation} 
The conditional prior for the waveform parameters of each binary can be further decomposed with the product rule and the fact that the intrinsic and extrinsic parameters are independent
\begin{equation}
\begin{split}
p(\vec\theta_i | \vec E, \mathcal{H},\mathcal{I}) =& p(m_{1i}, m_{2i} | \vec E, \mathcal{H},\mathcal{I}) p(\tilde\Lambda_i | m_{1i}, m_{2i}, \vec E, \mathcal{H},\mathcal{I})\\
&\times p(\theta_{{\rm ex,}i} | \mathcal{H}, \mathcal{I}).
\end{split}
\end{equation}
Likewise, the likelihood for the $n$ independent BNS observations is
\begin{equation}
p(D | \vec E,\vec\theta_1,\dots,\vec\theta_n, \mathcal{H},\mathcal{I}) = \prod_{i=1}^n p(d_i | \vec\theta_i, \mathcal{H},\mathcal{I}),
\end{equation}
and we have used the fact that the likelihood only depends on the waveform parameters to write $p(\vec d_i | \vec\theta_i, \vec E,\mathcal{H},\mathcal{I}) = p(\vec d_i | \vec\theta_i,\mathcal{H},\mathcal{I})$. (The waveform signal depends on $\vec E$ only through $\tilde\Lambda_i$ which is already included as a waveform parameter.)

The marginalized PDF (Eq.~\eqref{eq:margEOS}) is now
\begin{equation}
\begin{split}
p(\vec E | D,\mathcal{H},\mathcal{I}) = &\frac{1}{p(D | \mathcal{H},\mathcal{I})} 
\int d\vec\theta_{{\rm in,}1} \dots d\vec\theta_{{\rm in,}n} \\
& \times p(\vec E | \mathcal{H},\mathcal{I}) \prod_{i=1}^n \left[p(m_{1i}, m_{2i} | \vec E, \mathcal{H},\mathcal{I}) \right.\\
& \left.\times p(\tilde\Lambda_i | m_{1i}, m_{2i}, \vec E, \mathcal{H},\mathcal{I}) L(\vec d_i; \vec\theta_{{\rm in,}i}, \mathcal{H}, \mathcal{I}) \right],
\end{split}
\end{equation}
where we have defined the quasilikelihood for the intrinsic parameters as
\begin{equation}
\label{eq:quasilike}
L(\vec d_i; \vec\theta_{{\rm in,}i}, \mathcal{H}, \mathcal{I}) = \int d\vec\theta_{{\rm ex},i} p(\vec\theta_{{\rm ex,}i} | \mathcal{H}, \mathcal{I}) p(\vec d_i | \vec\theta_i,\mathcal{H},\mathcal{I}).
\end{equation}
Because $\tilde\Lambda_i$ is a deterministic function of $m_{1i}$, $m_{2i}$ and the EOS parameters,
\begin{equation}
\label{eq:lambdaeosrelation}
p(\tilde\Lambda_i | m_{1i}, m_{2i}, \vec E, \mathcal{H},\mathcal{I}) = \delta(\tilde\Lambda_i - \tilde\Lambda(m_{1i}, m_{2i}, \vec E)).
\end{equation}
The marginalized PDF finally becomes
\begin{equation}
\label{eq:margEOS2}
\begin{split}
p(\vec E | D,\mathcal{H},\mathcal{I}) =& \frac{1}{p(D | \mathcal{H},\mathcal{I})} \int dm_{11} dm_{21} \dots dm_{1n} dm_{2n} \\
&\times p(\vec E | \mathcal{H},\mathcal{I}) \prod_{i=1}^n \left[ p(m_{1i}, m_{2i} | \vec E, \mathcal{H},\mathcal{I}) \right.\\
& \left. \times L(\vec d_i; \vec\theta_{{\rm in,}i}, \mathcal{H}, \mathcal{I})|_{\tilde\Lambda_i = \tilde\Lambda(m_{1i}, m_{2i}, \vec E)} \right].
\end{split}
\end{equation}
The problem has now been reduced to computing the quasilikelihood (Eq.~\eqref{eq:quasilike}) for each BNS event, then computing Eq.~\eqref{eq:margEOS2}.

\subsection{Likelihood and signal to noise ratio}

The final ingredient we need to evaluate the marginalized PDF is an expression for the likelihood $p(\vec d_i | \vec\theta_i,\mathcal{H},\mathcal{I})$ for each GW event\footnote{In the following subsections, when we discuss the likelihood for individual GW events, we omit the event index $i$ for brevity.}. In this paper we assume that each detector in the network has stationary, Gaussian noise and that the noise between detectors is uncorrelated. This means that the power spectral density (PSD) $S_n(f)^a$ of the noise $n^a(t)$ in detector $a$ is
\begin{equation}
\langle \tilde n^a(f) \tilde n^{a*}(f') \rangle = \frac{1}{2} \delta (f-f') S_n(f)^a,
\end{equation}
where $\tilde n^a(f)$ is the Fourier transform of the noise of detector $a$, and $\langle\cdot\rangle$ represents an ensemble average. For a GW event with true parameters $\hat\theta$, resulting in the GW signal $h^a(t; \hat\theta)$, the data stream of detector $a$ will be
\begin{equation}
d^a(t) = n^a(t) + h^a(t; \hat\theta).
\end{equation}

For stationary, Gaussian noise, it is well known that the probability of obtaining the noise time series $n(t)$ is
\begin{equation}
p_n[n(t)] \propto e^{-(n, n)/2},
\end{equation}
where $(a, b)$ is the usual inner product between two time series $a(t)$ and $b(t)$ weighted by the PSD
\begin{equation}
(a, b)=4{\rm Re} \int_0^\infty \frac{\tilde a(f) \tilde b^*(f)}{S_n(f)}\,df.
\end{equation}
If we have a GW model $m(t;\vec\theta)$ that approximates the true signal $h(t;\vec\theta)$, then the likelihood of obtaining the data $d^a(t)$ given the GW model is
\begin{align}
p(d^a | \vec\theta, \mathcal{H}, \mathcal{I}) &= p_n[d^a(t)-m(t; \vec\theta)] \nonumber\\
& \propto e^{-(d^a - m(\vec\theta), d^a - m(\vec\theta))/2}.
\end{align}
If the model $m$ differs from the true signal $h$, a systematic error will be introduced in the recovered waveform parameters. For a network of independent detectors described by the set of time series $\vec d(t)$, the likelihood is then
\begin{equation}
p(\vec d | \vec\theta, \mathcal{H}, \mathcal{I}) = \prod_a p(d^a | \vec\theta, \mathcal{H}, \mathcal{I}).
\end{equation}

The observed signal to noise ratio (SNR) for the data $d^a(t)$ from detector $a$ is a Gaussian random variable given by
\begin{equation}
\rho^a = \frac{(d^a, m)}{\sqrt{(m, m)}}.
\end{equation}
The SNR for a network of detectors is then $\rho_{\rm net} = \sqrt{\sum_a(\rho^a)^2}$.  This is a measure of the relative power a given GW signal produces in a network of detectors.

\subsection{Averaged likelihood}
\label{sec:averagelike}


In order to determine the characteristic ability of GW detectors to measure EOS parameters, we will want to average our results for Eq.~\eqref{eq:margEOS2} over many realizations of the detector noise. Ref.~\cite{NissankeHolzHughes2010} defined the averaged likelihood as the geometric mean of an ensemble of $M$ identical detectors measuring the same GW event
\begin{equation}
p_{\rm ave}(d^a | \vec\theta, \mathcal{H}, \mathcal{I}) = \left[\prod_{k=1}^M p(d_k^a | \vec\theta, \mathcal{H}, \mathcal{I})\right]^{1/M}.
\end{equation}
Taking $M$ to be large, they found that this is equivalent to setting the noise $n^a(t)$ used to generate the data $d^a(t)$ equal to zero, but still using the characteristic noise PSD in the expression for the likelihood:
\begin{equation}
\label{eq:averagedlike}
p_{\rm ave}(d^a | \vec\theta, \mathcal{H}, \mathcal{I}) \propto e^{-\frac{1}{2}(m^a(\vec\theta) - h^a(\hat\theta) | m^a(\vec\theta) - h^a(\hat\theta))}.
\end{equation}
This zero-noise likelihood will allow us to examine characteristic measurement uncertainties in the EOS parameters independent of the particular noise realization. It will also allow us in Section~VI to separate the effects of systematic errors due to uncertainties in the waveform model from effects due to individual noise realizations.

\subsection{Implementation}

We evaluate Eq.~\eqref{eq:margEOS2} for the marginalized PDF in a two-step procedure, similar to that described by Steiner {\it et al.}~\cite{SteinerLattimerBrown2010}, where we first evaluate the quasilikelihood (Eq.~\eqref{eq:quasilike}) for each of the $n$ BNS systems, then evaluate Eq.~\eqref{eq:margEOS2}. In the first step where we evaluate the quasilikelihood $L(\vec d; \vec\theta_{\rm in}, \mathcal{H}, \mathcal{I})$, we use the MCMC sampler \texttt{LALInferenceMCMC} included in the LSC Algorithm Library~\cite{lal} as implemented in Ref.~\cite{WadeCreightonOchsner2014} and described in more detail in Ref.~\cite{VeitchRaymondFarr2014}. The prior for the extrinsic parameters $p(\vec\theta_{\rm ex} | \mathcal{H}, \mathcal{I})$ is given in Section~IIC of Ref.~\cite{VeitchRaymondFarr2014} with the additional uniform prior on $\delta\tilde\Lambda$ of $-500 \le \delta\tilde\Lambda \le 500$. The priors for the intrinsic parameters of interest $\{m_{1}, m_{2}, \tilde\Lambda\}$ are not contained in the quasilikelihood, so we are required to use flat priors. In most cases we used $1M_\odot \le m_2 \le m_1 \le 30M_\odot$ as in Ref.~\cite{VeitchRaymondFarr2014} and $0 \le \tilde\Lambda \le 3000$ as in Ref.~\cite{WadeCreightonOchsner2014}. However, when we injected BNS systems with component masses of $1M_\odot$ in Section~\ref{sec:varymass}, we used $0.5M_\odot \le m_2 \le m_1 \le 30M_\odot$ and $0 \le \tilde\Lambda \le 5000$ so that the posterior was minimally affected by the prior. For the likelihood $p(\vec d | \vec\theta, \mathcal{H}, \mathcal{I})$, we always use the TaylorF2 waveform as our GW model because it is the fastest to generate. After running the MCMC sampler, the marginalized distribution $L(\vec d; \vec\theta_{\rm in}, \mathcal{H}, \mathcal{I})$ is evaluated from the chain of $\{m_{1}, m_{2}, \tilde\Lambda\}$ samples with a Gaussian kernel density estimator.

In the second step, we sample the $4+2n$ parameter integrand of Eq.~\eqref{eq:margEOS2} using the affine-invarient ensemble sampler \texttt{emcee}~\cite{ForemanMackeyHogg2013}. For the prior on the EOS parameters $p(\vec E | \mathcal{H},\mathcal{I})$, we use uniform distributions with boundaries as described in Section~\ref{sec:eosparam} and additional boundaries from the $1.93M_\odot$ observation and causality constraints as described in Sections~\ref{sec:eosconstraints} and~\ref{sec:eosparam}. For the prior on the masses $p(m_{1i}, m_{2i} | \vec E, \mathcal{H},\mathcal{I})$, we use uniform distributions with $1M_\odot \le m_2 \le m_1 \le 3M_\odot$ or, when we examine $1M_\odot$ systems, $0.5M_\odot \le m_2 \le m_1 \le 3M_\odot$. The upper limit of $3M_\odot$ is sufficient to include NSs for all viable EOSs. In general, the dominant cost comes from evaluating $\tilde\Lambda(m_{1i}, m_{2i}, \vec E)$ for each BNS system at each iteration, but this is sped up by precomputing the five-parameter function $\Lambda(m, \vec E)$ on a grid and interpolating. We perform 10 runs with random initial parameters and test for convergence with the Gelman-Rubin diagnostic~\cite{GelmanRubin1992} before joining the samples. 
The marginalized distribution $p(\vec E|D, \mathcal{H}, \mathcal{I})$ is then given by a histogram of the samples for $\vec E$.

Evaluating the quasilikelihood (Eq.~\eqref{eq:quasilike}) with \texttt{LALInferenceMCMC} for all $n$ events in a population is very computationally expensive, so in some cases we use the Fisher matrix approximation instead. In the large SNR limit, the difference $\Delta\vec\theta = \vec\theta - \hat\theta$ between the estimated parameters $\vec\theta$ and the true parameters $\hat\theta$ of the binary system obeys a Gaussian distribution~\cite{FinnChernoff1993}. Specifically, for $N$ parameters, the likelihood is
\begin{equation}
\label{eq:gaussianlike}
p(d^a | \vec\theta,\mathcal{H},\mathcal{I}) = \frac{1}{\sqrt{(2\pi)^N \mathrm{det}(\Sigma_{ij})}} e^{-\frac{1}{2} \Sigma^{-1}_{ij} \Delta\theta_i \Delta\theta_j},
\end{equation}
where $\Sigma_{ij}$ is the covariance matrix, and it is given in terms of the Fisher matrix
\begin{equation}
\Gamma_{ij} = (\partial_i h(\hat\theta), \partial_j h(\hat\theta))
\end{equation}
by the relation $\Sigma_{ij}  = \Gamma^{-1}_{ij}$. In the large SNR limit, $\hat\theta$ will be approximately given by the maximum likelihood. When we use the Fisher matrix approximation we will use the mass variables $\ln(\mathcal{M})$ and $\ln(\eta)$ instead of $m_1$ and $m_2$ and flat priors for $\ln(\mathcal{M})$ and $\ln(\eta)$. The quasilikelihood (Eq.~\eqref{eq:quasilike}) marginalized over the extrinsic parameters is simply given by the submatrix of $\Sigma_{ij}$ containing the intrinsic parameters $\{\ln(\mathcal{M}), \ln(\eta), \tilde\Lambda\}$. However, when we estimate the EOS parameters in Eq.~\eqref{eq:margEOS2}, we always sample the posterior with \texttt{emcee}.

\section{Results}
\label{sec:results}

In this Section we characterize the ability of the aLIGO--aVirgo network to measure the EOS from a population of BNS inspiral events. For the two aLIGO detectors, we use the zero detuned-high laser power (broadband) PSD~\cite{LIGOnoise} which represents the design sensitivity that may be achieved for the aLIGO detectors by 2019~\cite{AasiAbadieAbbott2013}. For aVIRGO, which has higher high-frequency noise, we use the PSD fit from Ref.~\cite{ManzottiDietz2012} that may be achieved by 2021~\cite{AasiAbadieAbbott2013}. We also use the TaylorF2 waveform as both the injected GW signal $h(t; \hat\theta)$ and the GW model $m(t; \vec\theta)$ used for parameter estimation.

\subsection{Baseline BNS population}
\label{sec:baseline}

We start with a BNS population that has a realistic distribution of masses, number of events, and a moderate EOS, then later examine how these choices effect the results. We sample our population as follows:
\begin{enumerate}
\item \textit{Masses.} NSs in most BNS systems are thought to undergo little accretion after their formation, and are therefore found to be in the relatively narrow mass range characteristic of non-accreting NSs. The 10 currently known BNS systems~\cite{Lattimer2012} have most likely NS masses in the range $1.04M_\odot$--$1.53M_\odot$ (some of these have significant uncertainties) and are shown in Fig.~\ref{fig:rlofm}. \"Ozel {\it et al.}~\cite{OzelPsaltisNarayan2012}, for example, modeled the mass distribution of the known BNS systems as a Gaussian and found the most likely values for the mean and standard deviation to be $1.33M_\odot$ and $0.06M_\odot$ respectively. They also found that the mass ratios of the known BNS systems are consistent with each NS being drawn from this distribution independent of its companion. For simplicity, we draw the mass of each NS independent of its companion from a uniform distribution between $1.2M_\odot$ and $1.6M_\odot$. 

\item \textit{Events.} Significant uncertainty exists in the BNS inspiral event rate. Ref.~\cite{LIGORate2010} compiled rate estimates from several population-synthesis models and observations (Table 6 of Ref.~\cite{LIGORate2010}), and summarized the results as follows: a lower 95\% confidence bound of 1 event per Milky Way Equivalent Galaxy per Myr (MWEG$^{-1}$ Myr$^{-1}$), a most likely ``realistic'' value of 100~MWEG$^{-1}$ Myr$^{-1}$, and an upper 95\% confidence bound of 1000~MWEG$^{-1}$ Myr$^{-1}$. This corresponds to GW detection rates of 0.4~yr$^{-1}$, 40~yr$^{-1}$, and 400~yr$^{-1}$ respectively for a single aLIGO interferometer, using the broadband PSD with a threshold $\text{SNR} > 8$, averaging over sky location and orientation, with NS masses of $1.4M_\odot$, and a density of 0.0116~MWEG~Mpc$^{-3}$~\cite{LIGORate2010}. 

We simulated a year of GW data using the ``realistic'' event rate above. Specifically, we calculated the event rate in a volume large enough to contain all detectable BNS events. Because inspiral events are a Poisson process, we sampled the actual number of events in a year from a Poisson distribution with this rate. We then sampled the locations of these events uniformly in the volume, the orientations uniformly on a unit sphere, and the individual NS masses uniformly in $[1.2M_\odot, 1.6M_\odot]$. Of these systems $\sim$120 had a network SNR $\ge$ 8 and $\sim$30 had a network SNR $\ge$ 12. We performed parameter estimation for the 20 loudest (highest SNR) sources with network SNR that ranged from 63.7 down to 13.6, integrating between the GW frequencies $f_{\rm low} = 30$~Hz and $f_{\rm ISCO}$.

\item \textit{EOS.} We used the piecewise-polytrope fit to the MPA1 EOS given in Table~\ref{tab:eosfit} which has a radius and maximum mass roughly in the middle of the range, then calculated the corresponding tidal parameter for each sampled NS. Using the fit instead of the tabulated EOS separates the systematic error due to the inexact EOS fit from the analysis of statistical errors presented here, and we leave the discussion of these systematic errors to Section~\ref{sec:systematicEOS}.

\end{enumerate}

\begin{figure*}[!htb]
\begin{center}
\includegraphics[width=3.2in]{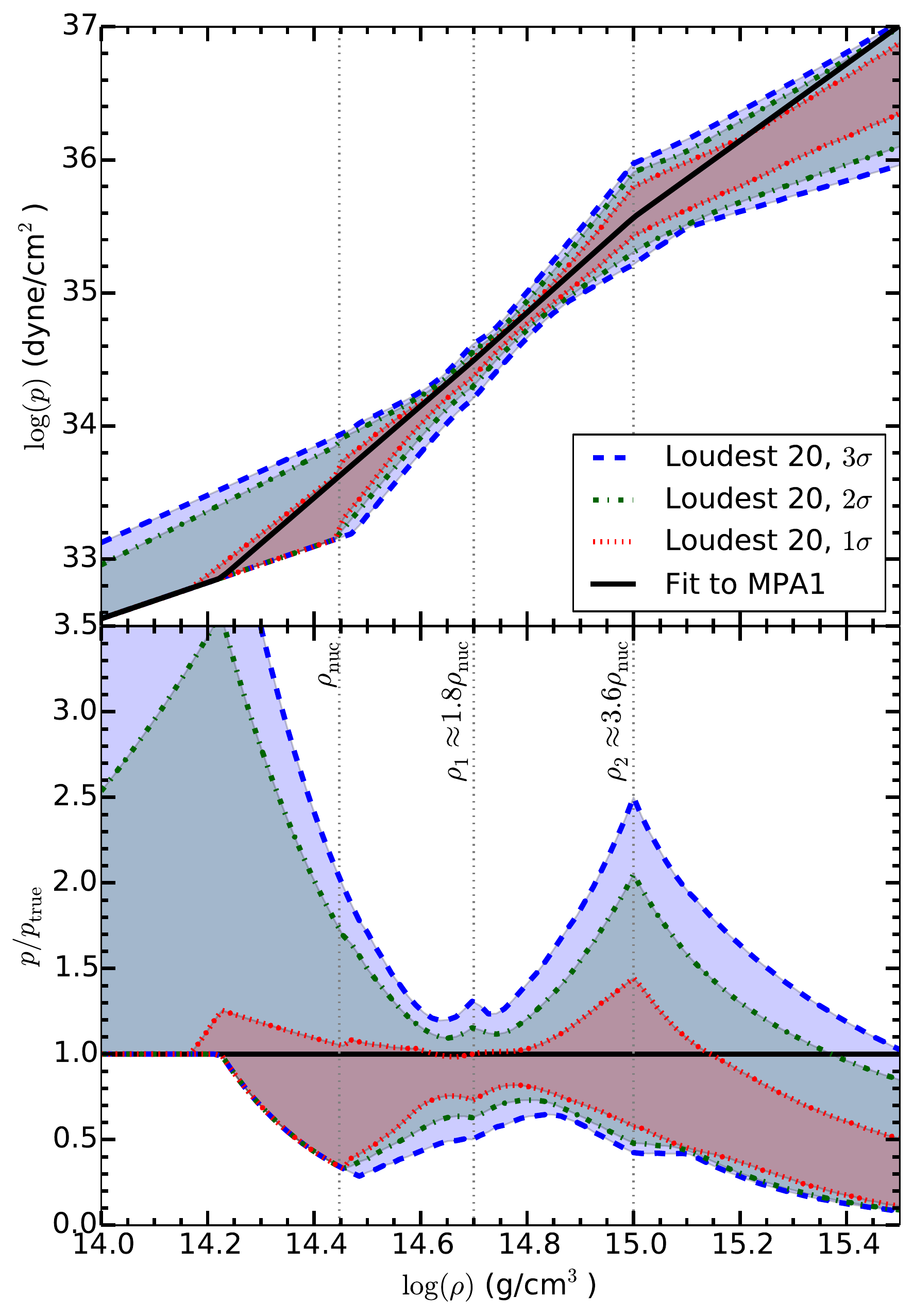}
\includegraphics[width=3.2in]{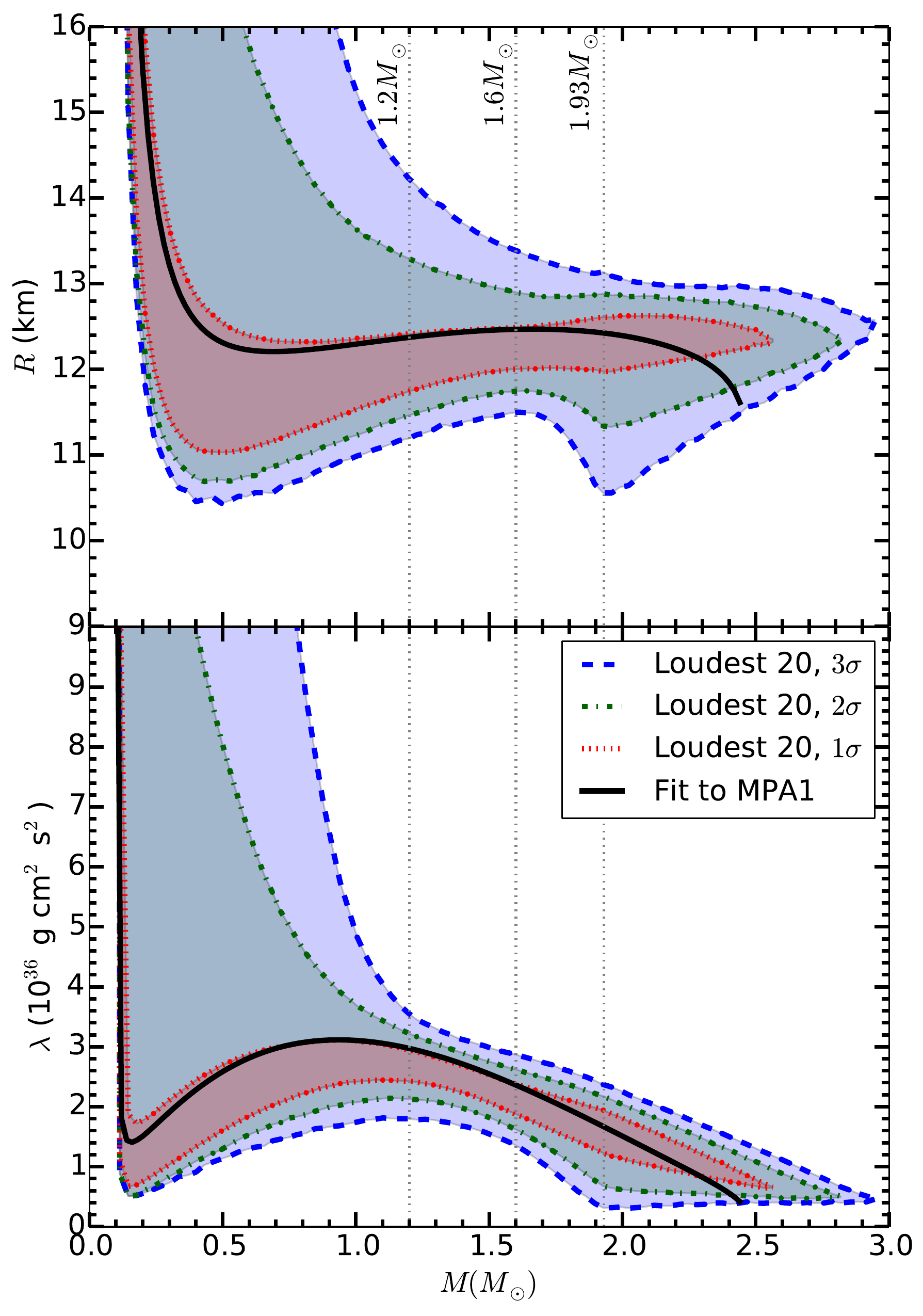}
\caption{Uncertainty in the recovered EOS, radius, and tidal deformability for the three-detector aLIGO--aVirgo network. Results are shown for the loudest 20 events with network SNR from 63.7 down to 13.6. The red, green, and blue regions represent the 68\% ($1\sigma$), 95\% ($2\sigma$), and 99.7\% ($3\sigma$) credible regions respectively. In the bottom left panel, $p_{\rm true}$ is the pressure of the ``true'' injected EOS, which in this case is the fit to the MPA1 EOS. In the right panels, the vertical line at $1.93M_\odot$ is the minimum mass, set by the prior, where some accepted EOS parameters do not produce a stable NS.}
\label{fig:mpa1fit}
\end{center}
\end{figure*}

Using zero-noise data described in Section~\ref{sec:averagelike}, we show in Fig.~\ref{fig:mpa1fit} the measurability of the EOS, radius, and tidal deformability for the loudest 20 events in our population. The contours represent the 68\% ($1\sigma$), 95\% ($2\sigma$), and 99.7\% ($3\sigma$) credible regions. In the left panel we plot the credible region for $\log(p)$ as well as for $p/p_{\rm true}$, where we call the piecewise-polytrope fit to the MPA1 EOS the ``true'' EOS. We generate these figures as follows: At the density $\rho$, we evaluate $\log[p(\rho, \vec E)]$ for each set of EOS parameters from the MCMC simulation. We then evaluate the credible interval at that density from the sampled $\log(p)$ values. The same is done for $p/p_{\rm true}$. In the top right panel, we generated the credible interval for each mass $M$ from the radii samples that were found by evaluating $R(M, \vec E)$ for each set of EOS parameters in the MCMC simulation. For masses greater than the $1.93M_\odot$ prior, some of the sampled EOS parameters do not allow for a stable NS. For those sampled EOS parameters, an object of that mass would lead to a black hole. In this case, the distribution of radii becomes bimodal, with a delta function at the Schwarzschild radius $2GM/c^2$ and weight proportional to the fraction of samples that do not allow for a stable NS. The credible interval then represents the fraction of MCMC samples that produce a NS in that radius interval \textit{or} a black hole. The bottom right panel shows the confidence intervals for the tidal deformability $\lambda$. For the samples that produce a black hole above $1.93M_\odot$, $\lambda=0$~\cite{BinningtonPoisson2009}.

Fig.~\ref{fig:mpa1fit} has sharp peaks in the fractional uncertainty $p/p_{\rm true}$ around the variable transition density $\rho_0 \lesssim \rho_{\rm nuc}$ and at the fixed transition densities $\rho_1$ and $\rho_2$ between the polytrope pieces. This is due to the choice of parameterized EOS model. Allowing the transition densities to be additional free parameters would likely smooth these features out. Indeed, the EOS fit in Steiner {\it et al.}~\cite{SteinerLattimerBrown2010} which included the transition densities as free parameters did not show such features in the results for $p/p_{\rm true}$. 

Above the transition density $\rho_2$, the results increasingly underpredict the pressure. This occurs because, although the MPA1 EOS is causal with $v_{\rm s,max}=0.994$ (Table~\ref{tab:eosfit}), the corresponding piecewise-polytrope fit overpredicts $v_{\rm s,max}$ by $\sim 5\%$ and is therefore acausal at high densities. However, the accepted MCMC samples are required to have $v_{\rm s,max} \leq c$, resulting in accepted samples corresponding to smaller pressures. 

The credible interval is largest at densities below $\sim \rho_{\rm nuc}$ and for corresponding low mass stars where the densities are lower. This results because the bulk of NS matter is above $\sim \rho_{\rm nuc}$, and we included minimal {\it a priori} information on how the core EOS joins onto the lower-density crust EOS. In contrast, Steiner {\it et al.}~\cite{SteinerLattimerBrown2010} parameterized the EOS around $\rho_{\rm nuc}$ in terms of the baryon density and proton fraction with 4 free parameters (Eq.~(33) of Ref.~\cite{SteinerLattimerBrown2010}), and this provides stronger {\it a priori} constraints on the behavior below $\sim \rho_{\rm nuc}$ (Fig.~8 of Ref.~\cite{SteinerLattimerBrown2010}). Overall, it is clear that in some density regions, a significant contribution to the credible interval comes from our choice of EOS parameterization which was not optimized for the purposes here, rather than from the sensitivity of the GW detectors. 

We also find that the error in the tidal deformability $\lambda$ is smallest in the mass interval $1.2M_\odot$--$1.6M_\odot$ where the BNS masses were drawn from. This is not surprising. However, $\lambda$ can still be measured with comparable accuracy for a much larger range of masses. This is in contrast to the results of Del Pozzo {\it et al.}~\cite{DelPozzoLiAgathos2013} that did not incorporate the additional information about the EOS presented in Section~\ref{sec:eosconstraints}.

\begin{figure}[!htb]
\begin{center}
\includegraphics[width=3.2in]{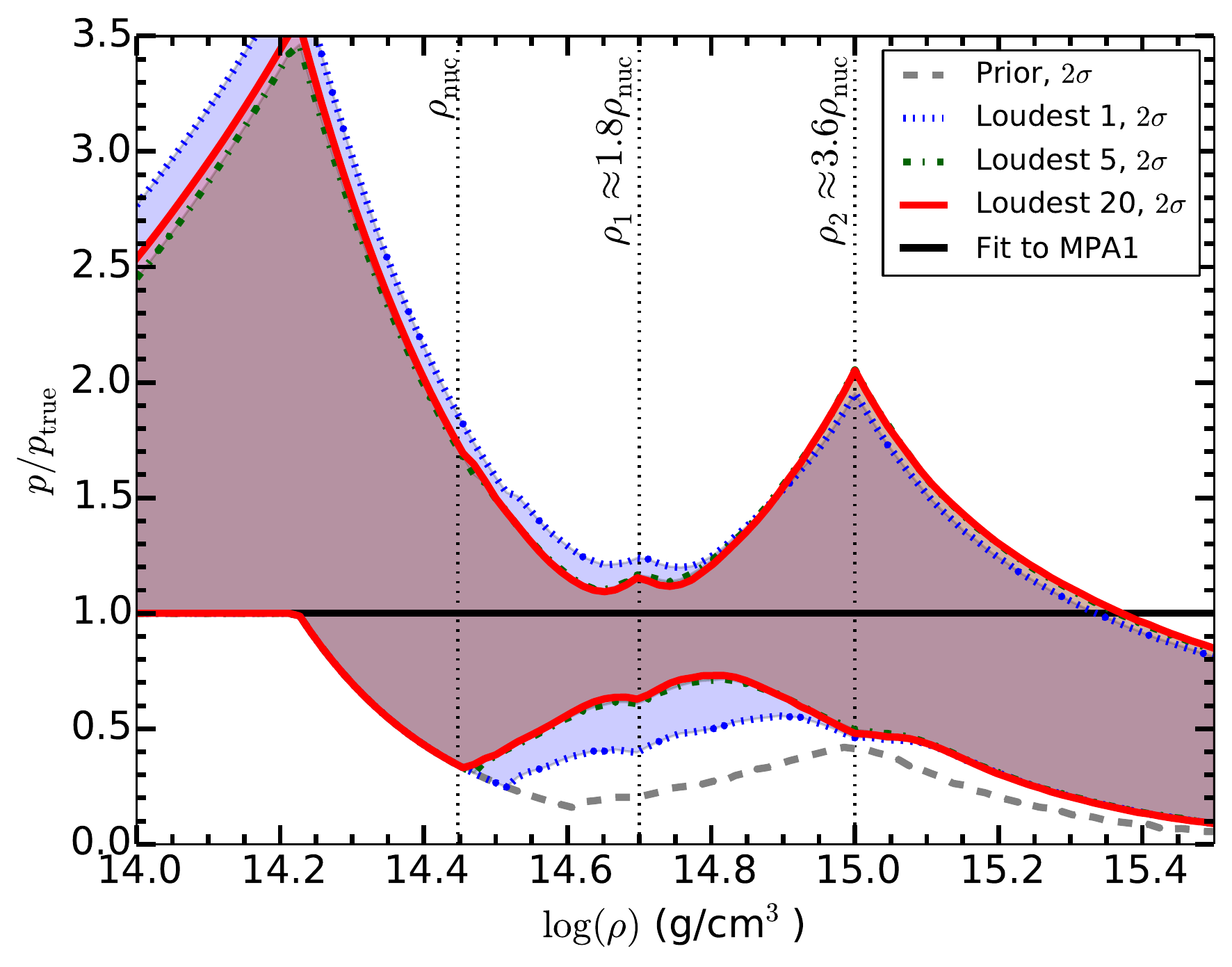}\\
\includegraphics[width=3.2in]{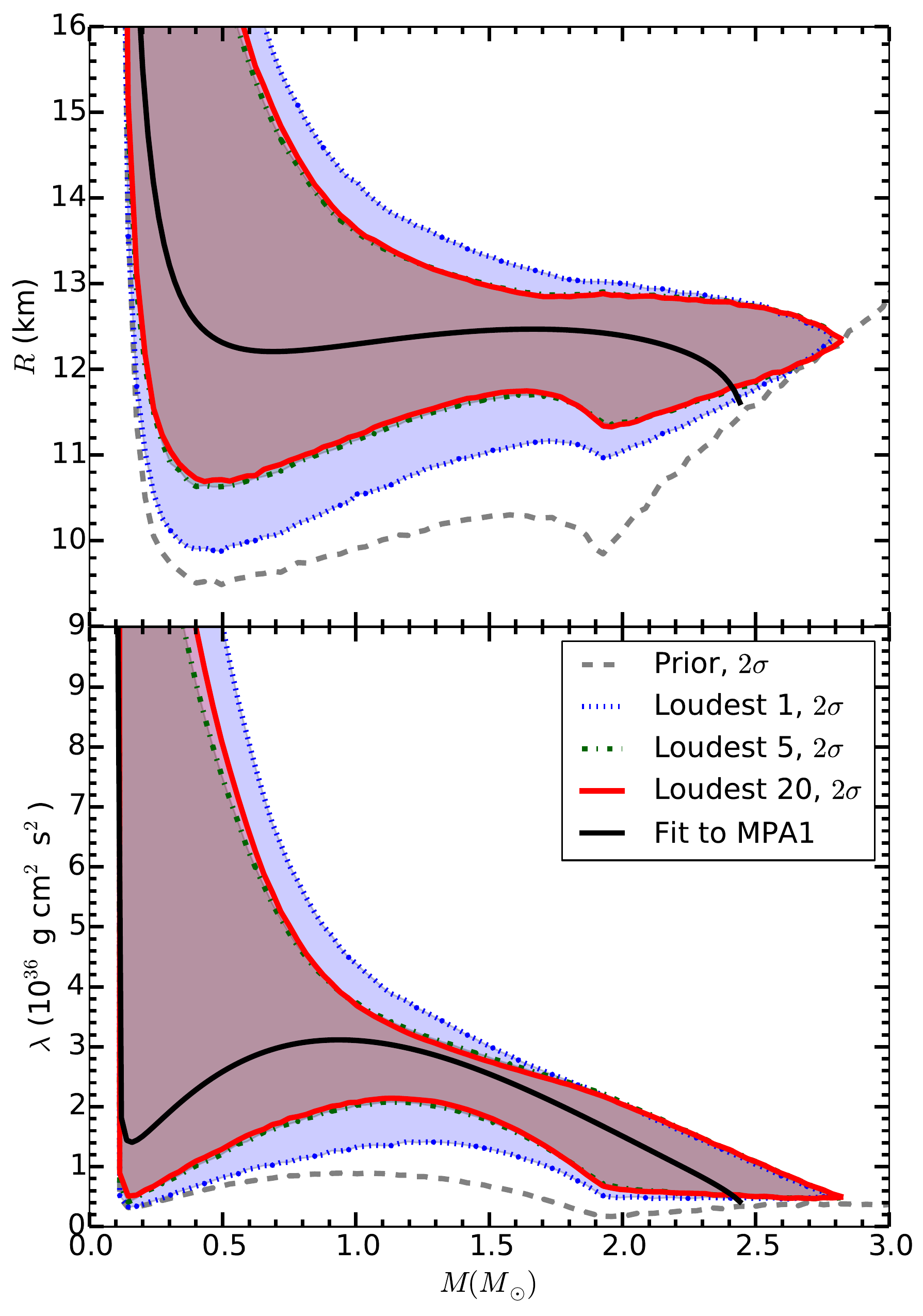}
\caption{Same BNS population as Fig.~\ref{fig:mpa1fit}. Contours represent the 95\% credible regions for the loudest 1, 5, and 20 inspiral events. Including quieter events from the population does not improve the results. Also shown is the lower limit on the 95\% credible region from just the maximum mass ($M_{\rm \max} \geq 1.93M_\odot$) and causality priors.}
\label{fig:mpa1fitofn}
\end{center}
\end{figure}

Finally, we show how the credible region depends on the number of events in Fig.~\ref{fig:mpa1fitofn}. The dashed gray curve represents the lower limit set by the priors without any BNS inspiral data. This lower limit for the radius of $\sim 10$~km above $1M_\odot$ is in mild tension with Guillot {\it et al.}~\cite{GuillotServillatWebb2013} who, combining data from observations of several NSs, found that the NS radius is 7.6--10.4km (90\% confidence). A similar lower bound from the maximum mass and causality constraints was found in Ref.~\cite{HebelerLattimerPethick2013} (dotted curve in the left panel of Fig.~11 of Ref.~\cite{HebelerLattimerPethick2013}). However, by softening the EOS so that $v_s = \sqrt{dp/d\epsilon} = c$ whenever their piecewise polytrope parameterization became acausal, they were able to weaken this constraint, and their lower limit on the radius is $\sim 1$~km less than the one presented here. The upper limit from the prior (not shown here) is a few times larger than the scale of the figures and is set mainly by the causality constraint. This is in contrast to Ref.~\cite{HebelerLattimerPethick2013}. Because they used strong assumptions about the EOS below $\sim \rho_{\rm nuc}$ from chiral effective field theory, they were able to place an upper limit on the NS radius of $\sim 15$~km. Because we use a much less restricted low-density EOS, the pressure at higher densities is allowed to have much larger values, resulting in larger radii.

When data from the BNS inspirals are included, the overwhelming majority of the information about the EOS is obtained from just the loudest 5 events in the population. After the loudest 5 events, including more events does not improve the measurability of the EOS parameters, radius, or tidal deformability.

\subsection{Highest known NS mass}

As discussed in Section~\ref{sec:eosconstraints}, the highest mass NSs with rigorous constraints are $\sim 2M_\odot$~\cite{DemorestPennucciRansom2010, AntoniadisFreireWex2013}, and we have used $1.93M_\odot$ as the lower bound on the maximum mass. However, there is also evidence for NSs with higher masses. In particular, a class of pulsars known as black-widows irradiate their companions and generate outflows that are accreted onto the pulsar, significantly increasing the pulsar's mass. Using spectra to determine the radial velocity of the companion, PSR B1957+20 was found to have a mass of $2.40\pm0.12M_\odot$ after correcting for the anisotropic emission of the companion which causes the center of light to lie inward relative to the center of mass. However, considering systematic uncertainties in the model they found a conservative lower limit of $1.66M_\odot$~\cite{VanKerkwijkBretonKulkarni2011}. Another black-widow pulsar, PSR J1311-3430, was found to have a mass of $2.68\pm0.14M_\odot$, but considering similar uncertainties, a lower limit of $2.1M_\odot$ was claimed~\cite{RomaniFilippenkoSilverman2012}. 

\begin{figure}[!htb]
\begin{center}
\includegraphics[width=3.2in]{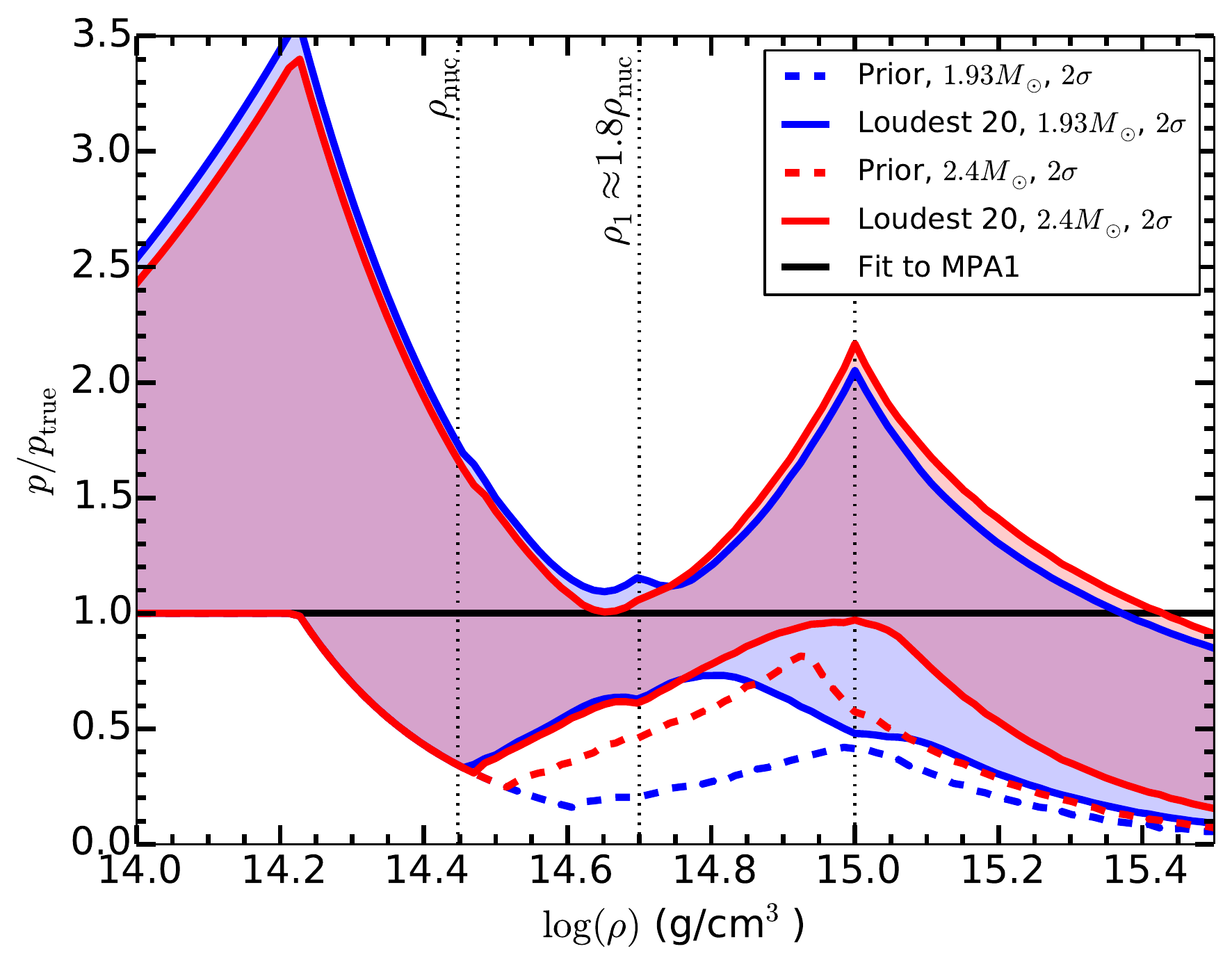}\\
\includegraphics[width=3.2in]{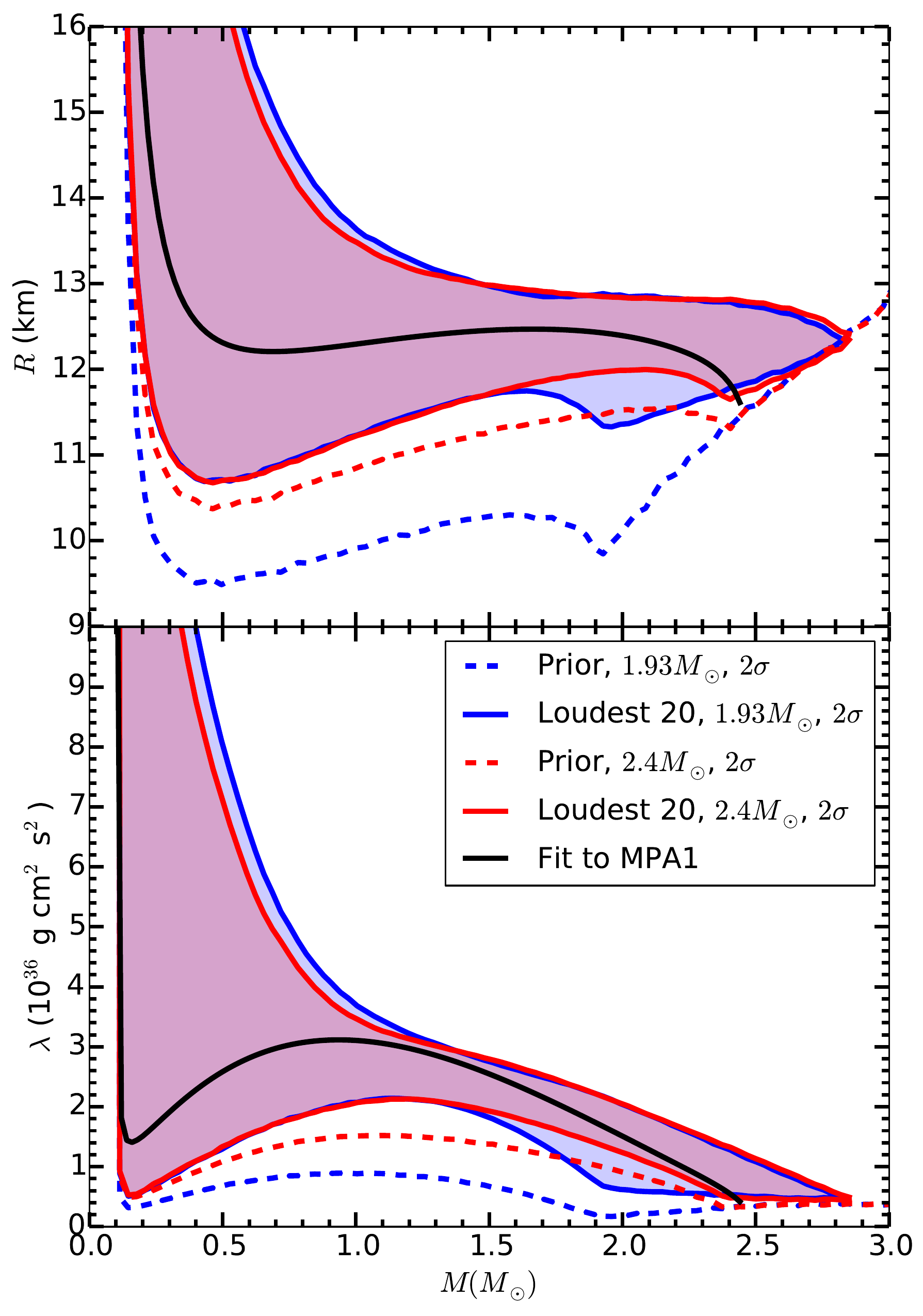}
\caption{Dependence of credible regions on the highest known NS mass. 95\% credible regions are shown for the same 20 BNS systems as in Fig.~\ref{fig:mpa1fit}. The maximum mass prior is set to $M_{\rm max} \geq 1.93M_\odot$ for the red shaded region and $M_{\rm max} \geq 2.4M_\odot$ for the blue shaded region. Also shown are the lower limits on the 95\% credible regions from just the maximum mass and causality priors, with $M_{\rm max} \geq 1.93M_\odot$ (red dashed) and $M_{\rm max} \geq 2.4M_\odot$ (blue dashed).}
\label{fig:varymmax}
\end{center}
\end{figure}

In Fig.~\ref{fig:varymmax} we demonstrate that the confirmation of a NS with a lower mass bound of $2.4M_\odot$, for example, would place a significantly tighter lower bound on the pressure for $\rho \gtrsim \rho_1$. The maximum mass of the MPA1 EOS is $2.461M_\odot$ ($2.457M_\odot$ for the piecewise-polytrope fit), so this EOS would almost be ruled out. In contrast, the upper bound on the pressure above $\sim\rho_2=10^{15}$~g/cm$^3$ in Fig.~\ref{fig:varymmax} comes from the causality requirement. (Recall from Fig.~\ref{fig:mvs3d}, the causality constraint restricts large values of $\Gamma_3$ and partially restricts large values of $\Gamma_2$.) Higher mass measurements will therefore not decrease the upper limit on the pressure at the highest NS densities. Similarly, observations of higher mass NSs place tighter lower bounds on the radius and tidal deformability at high masses, but do not improve the upper bound.

\subsection{Distribution of BNS masses}
\label{sec:varymass}

The density profile in a NS depends on its mass, with more massive stars consisting of denser matter. We thus expect low mass stars to better estimate the lower density EOS, and higher mass stars to better estimate the higher density EOS. We generate four BNS populations with different mass distributions, then examine how the error in recovering the EOS depends on the NS masses. The first population is the same as above, using masses uniformly sampled in the range $1.2M_\odot$--$1.6M_\odot$. We also examine three additional populations where the NSs are either all $1.0M_\odot$, all $1.4M_\odot$, or all $1.8M_\odot$. In order to make a direct comparison between these populations, we hold all of the parameters fixed except for the masses, then adjust the tidal parameters of each system as determined by the fit to the MPA1 EOS. Additionally, we only examine the loudest 5 systems in each population which, as shown above, contain the majority of the EOS information. As a result, adjusting their masses in this range will not push these events above or below the detection threshold. 

\begin{figure}[!htb]
\begin{center}
\includegraphics[width=3.2in]{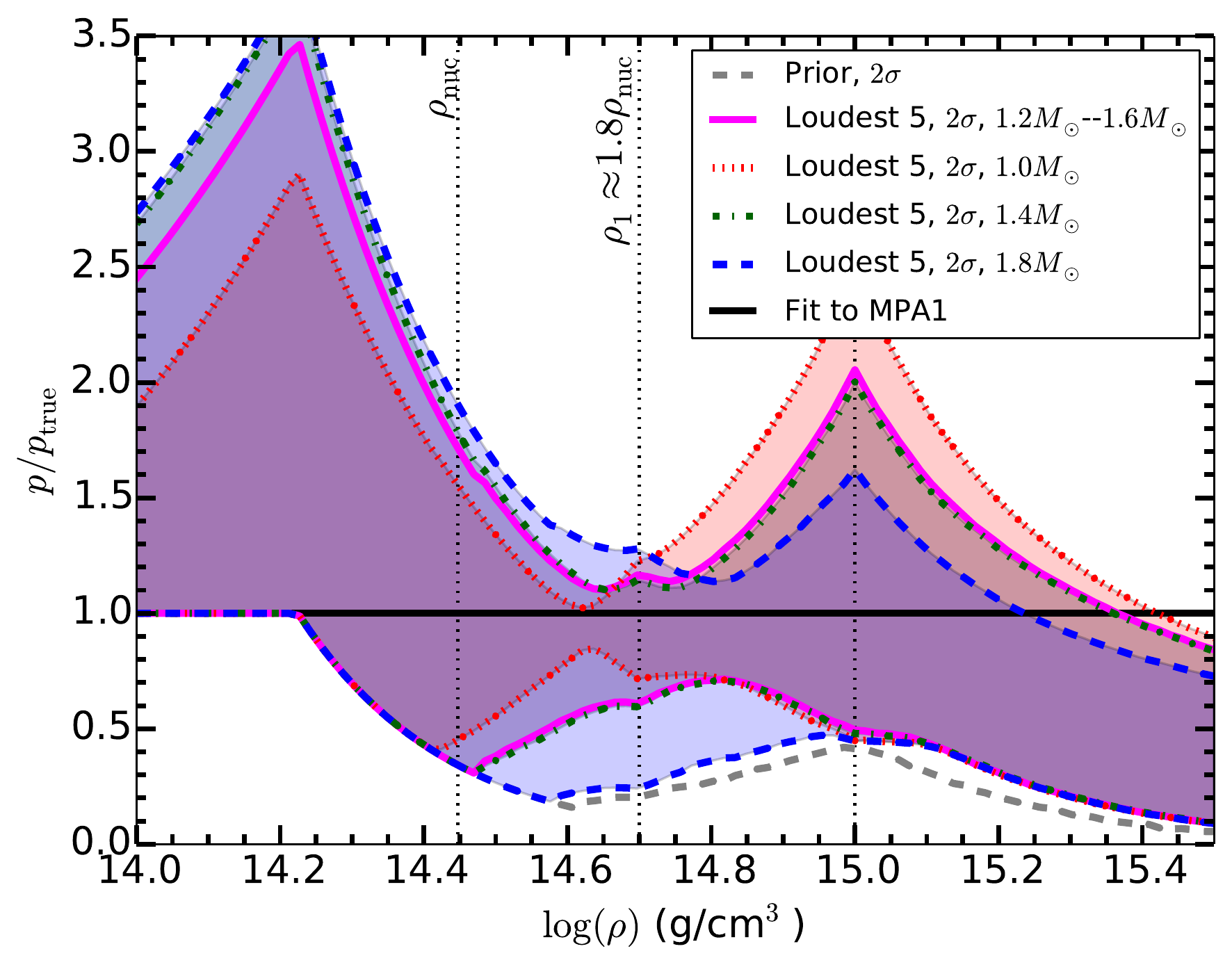}\\
\includegraphics[width=3.2in]{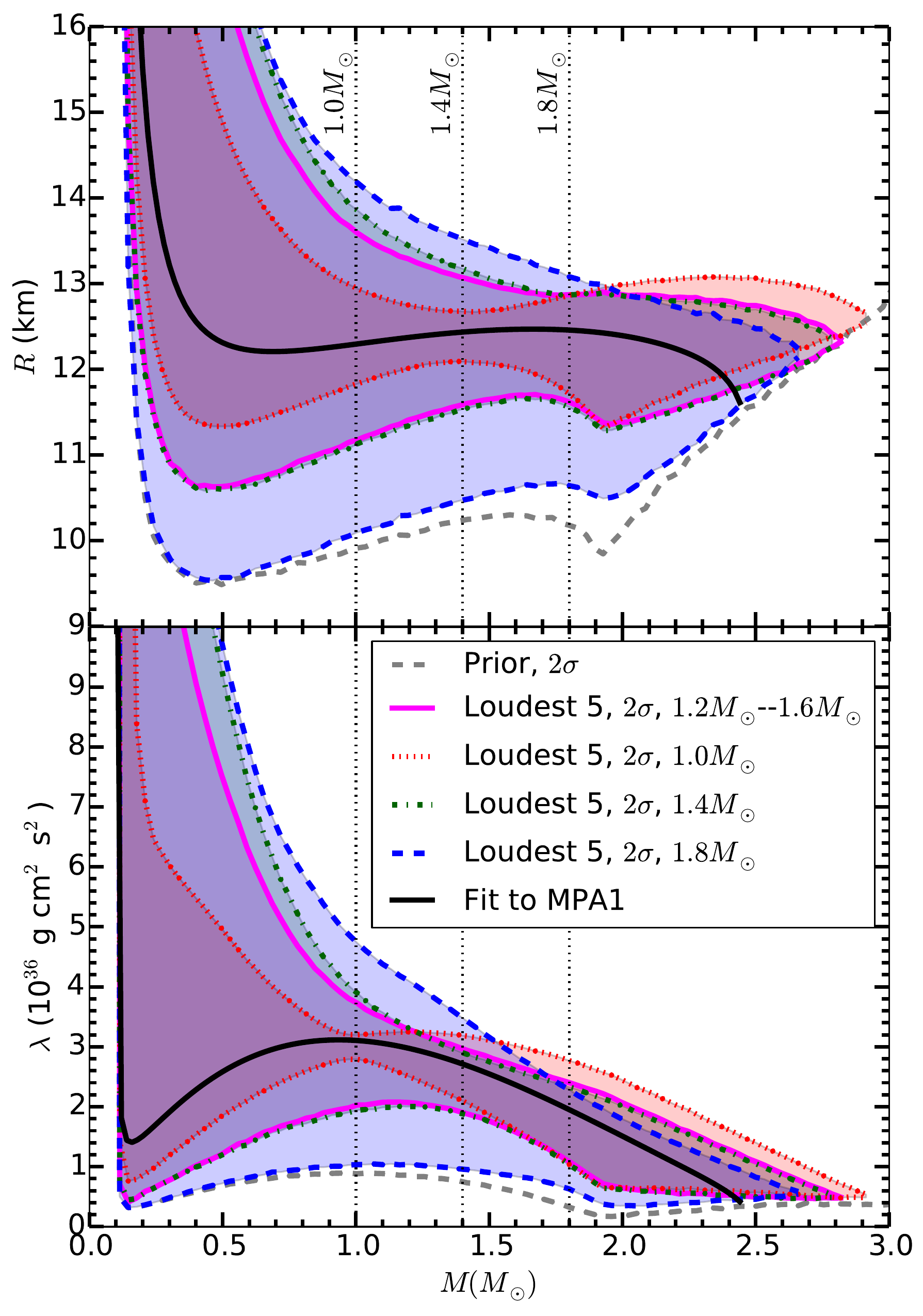}
\caption{Dependence of credible regions on the masses of NSs in the simulated BNS population. Results use the same BNS parameters as in Fig.~\ref{fig:mpa1fit}, except the masses are varied to be all $1.0M_\odot$ (green dot-dashed), all $1.4M_\odot$ (blue dashed), and all $1.8M_\odot$ (magenta). Also shown is the lower limit from just the maximum mass ($M_{\rm \max} \geq 1.93M_\odot$) and causality priors.}
\label{fig:popmass}
\end{center}
\end{figure}

In Fig.~\ref{fig:popmass}, we find that when all NSs have masses of either $1.0M_\odot$, $1.4M_\odot$, or $1.8M_\odot$, the uncertainty in pressure is smallest around $10^{14.6}$~g/cm$^3$, $10^{14.7}$~g/cm$^3$, or $10^{14.8}$~g/cm$^3$ respectively. We find similar results for the radius and tidal deformability; the location of the minimum uncertainty scales with the observed masses, and for the tidal deformability, the minima occur very close to the masses of the observed NSs. Interestingly, the results for masses fixed at $1.4M_\odot$ and for a uniform distribution from $1.2M_\odot$--$1.6M_\odot$ are almost identical, indicating that useful information can be found even if the range of observed masses is very small.

\subsection{Comparison of \texttt{LALInferenceMCMC} with Fisher matrix}
\label{sec:fisher}

In addition to the results found above by evaluating the quasilikelihood (Eq.~\eqref{eq:quasilike}) with \texttt{LALInferenceMCMC}, we also use the Fisher matrix approximation (Eq.~\eqref{eq:gaussianlike}) to evaluate the quasilikelihood. This approach is significantly faster but only accurate in the limit of large SNR signals. For the Fisher matrix approximation, we used a single detector with the broadband PSD, but scaled the amplitude such that the SNR was equal to the network SNR for the full aLIGO--aVirgo network. The waveform was identical to the one used with \texttt{LALInferenceMCMC}, except we did not use the term in the next-to-leading order tidal correction containing $\delta\tilde\Lambda$, and thus $\tilde\Lambda$ was our only tidal parameter. Because $\delta\tilde\Lambda$ has an insignificant impact on the waveform relative to $\tilde\Lambda$ and is unmeasurable as discussed in Ref.~\cite{WadeCreightonOchsner2014}, this is approximately equivalent to placing a reasonable flat prior on $\delta\tilde\Lambda$ then marginalizing over its values as is done in the \texttt{LALInferenceMCMC} calculation. Furthermore, we use flat priors for $\ln(\mathcal{M})$ and $\ln(\eta)$ in the Fisher matrix calculation instead of the flat priors for $m_1$ and $m_2$ used in the \texttt{LALInferenceMCMC} calculation.

\begin{figure}[!htb]
\begin{center}
\includegraphics[width=3.2in]{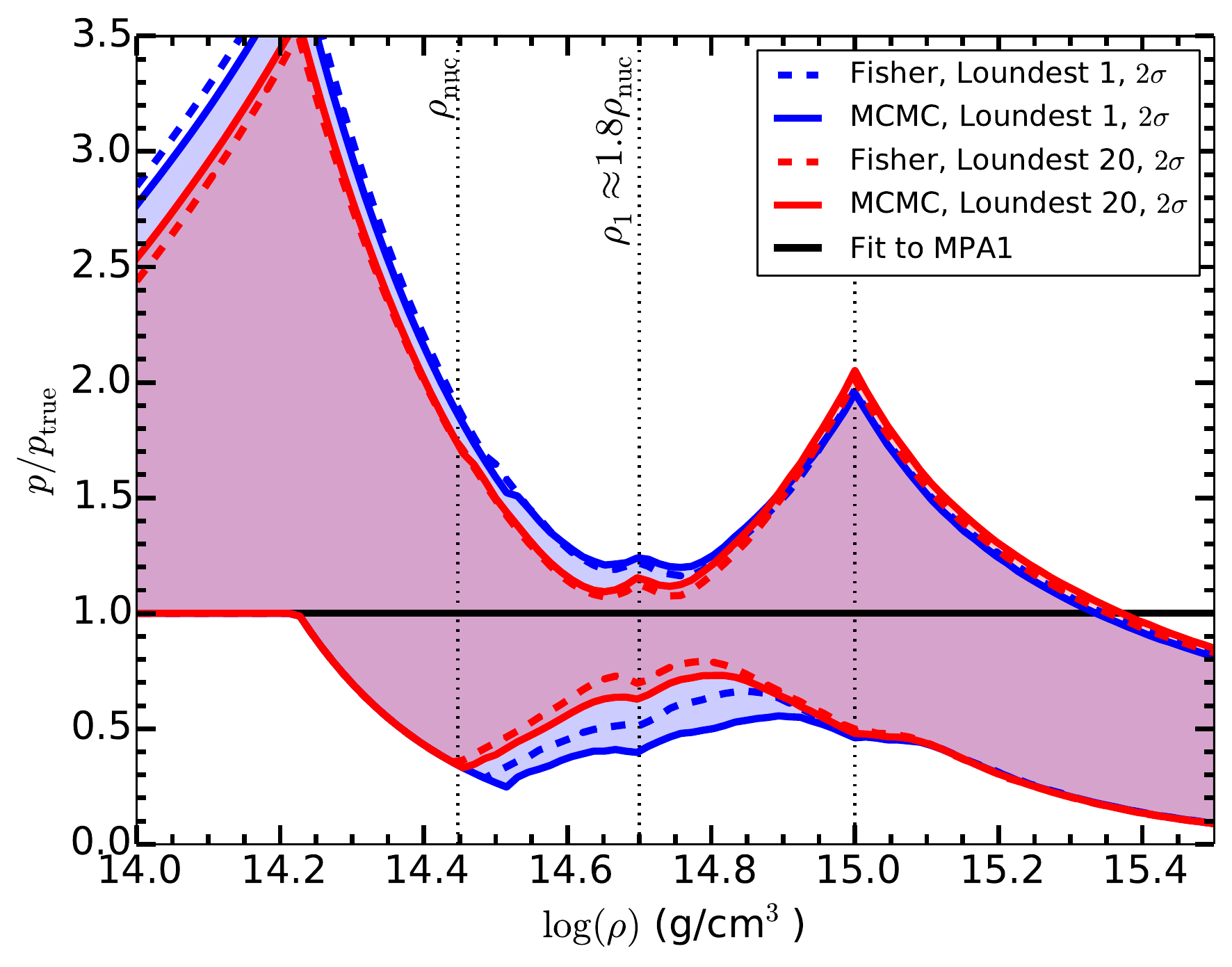}\\
\includegraphics[width=3.2in]{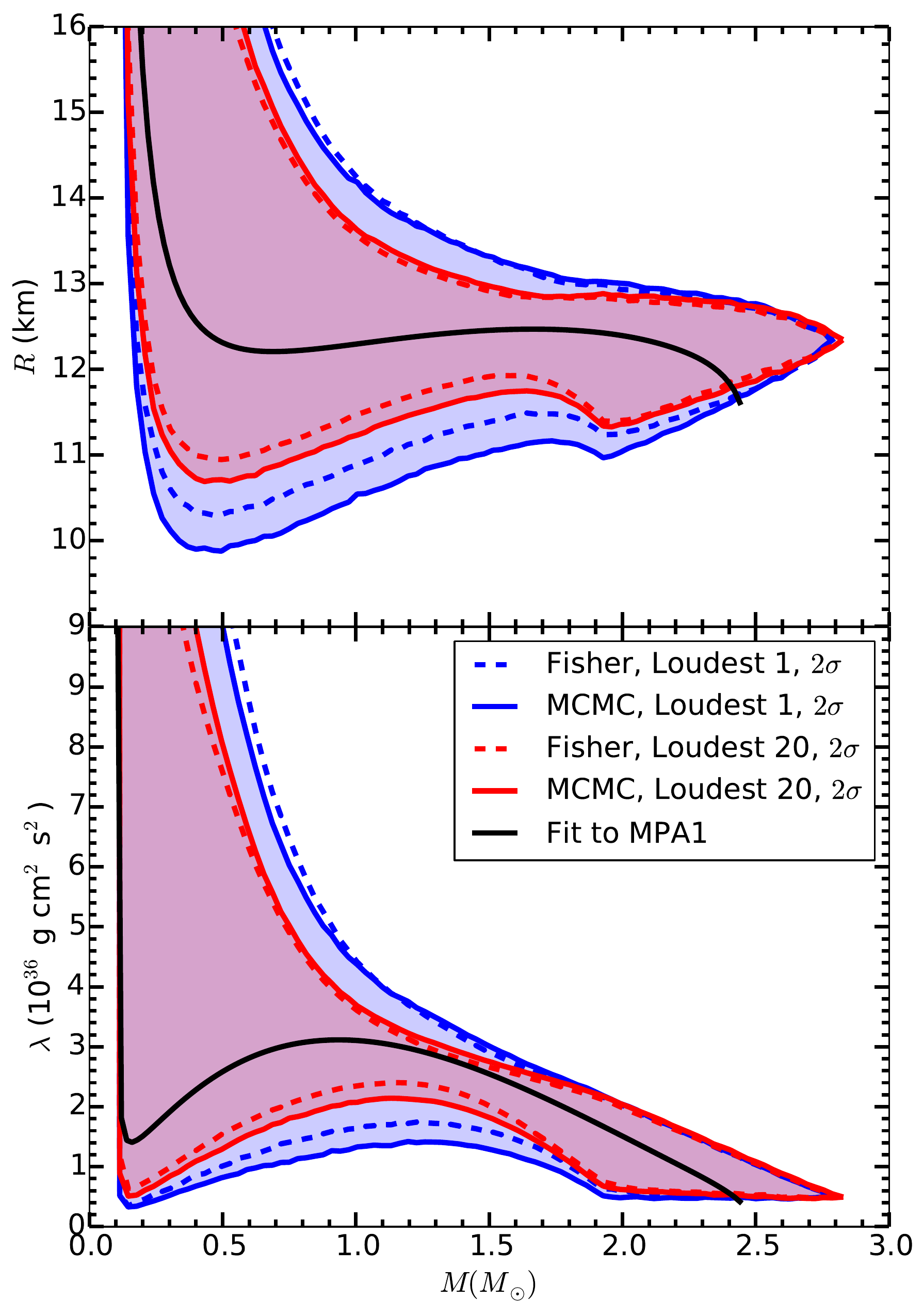}
\caption{Same BNS population as Fig.~\ref{fig:mpa1fit}. The SNR in the Fisher matrix calculation is set to match the network SNR of the \texttt{LALInferenceMCMC} calculation. Contours represent 95\% credible regions.}
\label{fig:fisher}
\end{center}
\end{figure}

In Fig.~\ref{fig:fisher}, we compare the results for calculating the quasilikelihood with \texttt{LALInferenceMCMC} and the Fisher matrix approximation.  We find that even with the differences listed above, the results are still comparable. The Fisher matrix approximation, however, slightly underestimates the uncertainty in the pressure, radius, and tidal deformability.

\subsection{Dependence on EOS model}

In this subsection we determine how the measurability of the EOS depends on the choice of ``true'' EOS. To do this, we use the same population of BNS systems as in Section~\ref{sec:baseline}, then vary the EOS and corresponding tidal deformability. Here we use the tabulated EOS models listed in Table~\ref{tab:eosfit} instead of the least-squares fit used above. This will allow us to examine systematic errors in Section~\ref{sec:systematicEOS} from the inexact EOS parameterization. For efficiency we also calculate the quasilikelihood with the Fisher matrix approximation which gives results consistent with \texttt{LALInferenceMCMC} as shown above.

\begin{figure*}[!htb]
\begin{center}
\includegraphics[width=6.4in]{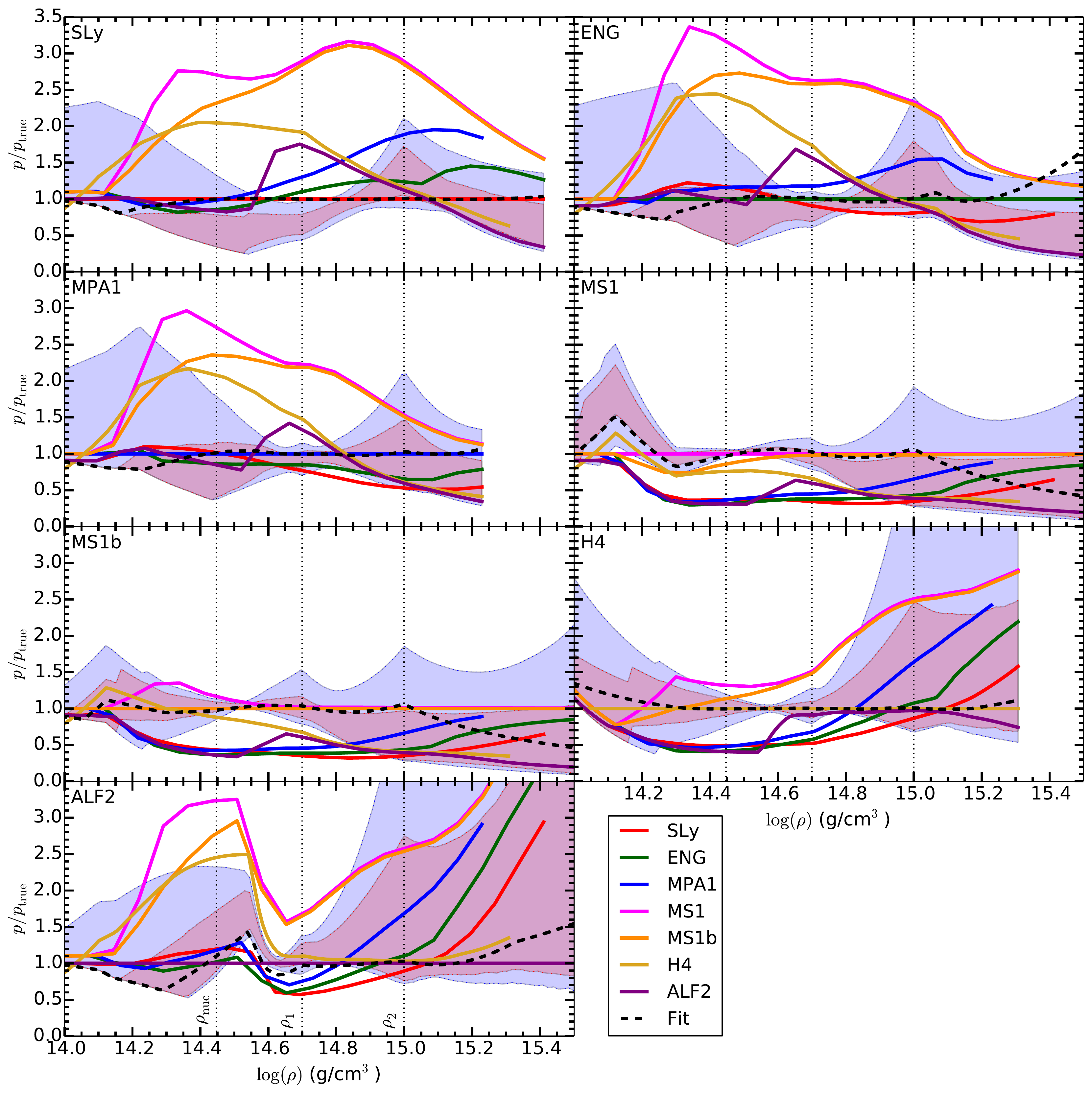}
\caption{Dependence of the EOS credible region on the ``true'' EOS for the 20 loudest BNS systems. Contours represent the 68\% and 95\% credible regions. The BNS parameters are the same as in Fig.~\ref{fig:mpa1fit} except for the choice of tabulated EOS. Results were calculated using the Fisher matrix approximation to the quasilikelihood, and the SNR in the Fisher matrix calculation was scaled to match the network SNR of the three-detector network as in Fig.~\ref{fig:fisher}. From left to right then top to bottom, the ``true'' EOSs are SLy, ENG, MPA1, MS1, MS1b, H4, and ALF2. Unlike the results above, the ``true'' EOSs used here are the tabulated EOSs, so the recovered EOSs also include the systematic bias from the inexact EOS parameterization. The corresponding best-fit piecewise polytropes are shown as dashed curves. For each panel, the other EOSs are also shown relative to the ``true'' tabulated EOS. The quantity $p/p_{\rm true}$ is not defined above the highest density in the EOS table, and for SLy, MPA1, and H4, the EOS table ends before the $10^{15.5}$~g/cm$^3$ right limit of each panel.}
\label{fig:perroralleos}
\end{center}
\end{figure*}

The uncertainty in the pressure for these EOSs is shown in Fig.~\ref{fig:perroralleos}, and there are a few features to note. First, if the true EOS has a maximum speed of sound below the central density of the maximum mass NS $v_{\rm s, max}$ that is close to $c$ (SLy, ENG, MPA1, MS1, MS1b), the causality constraint places a useful upper bound on the pressure estimate $p(\rho)/p_{\rm true}(\rho)$ at densities above $\gtrsim \rho_2$. However, for the H4 and ALF2 EOSs where $v_{\rm s, max} \sim 0.6$, the causality requirement only provides a weak constraint on the high-density EOS. Second, we note that softer EOSs (lower pressures) result in stars that are more easily compressed (smaller radii) and have higher densities. They will therefore probe higher densities. We find that for the softer EOSs SLy, ENG, and MPA1, the uncertainty in the pressure is minimized at densities $\rho \gtrsim \rho_1$, whereas for the stiffer EOSs MS1, MS1b, H4, and ALF2, the uncertainty in the pressure is minimized at densities $\rho \lesssim \rho_1$.

\begin{figure*}[!htb]
\begin{center}
\includegraphics[width=6.4in]{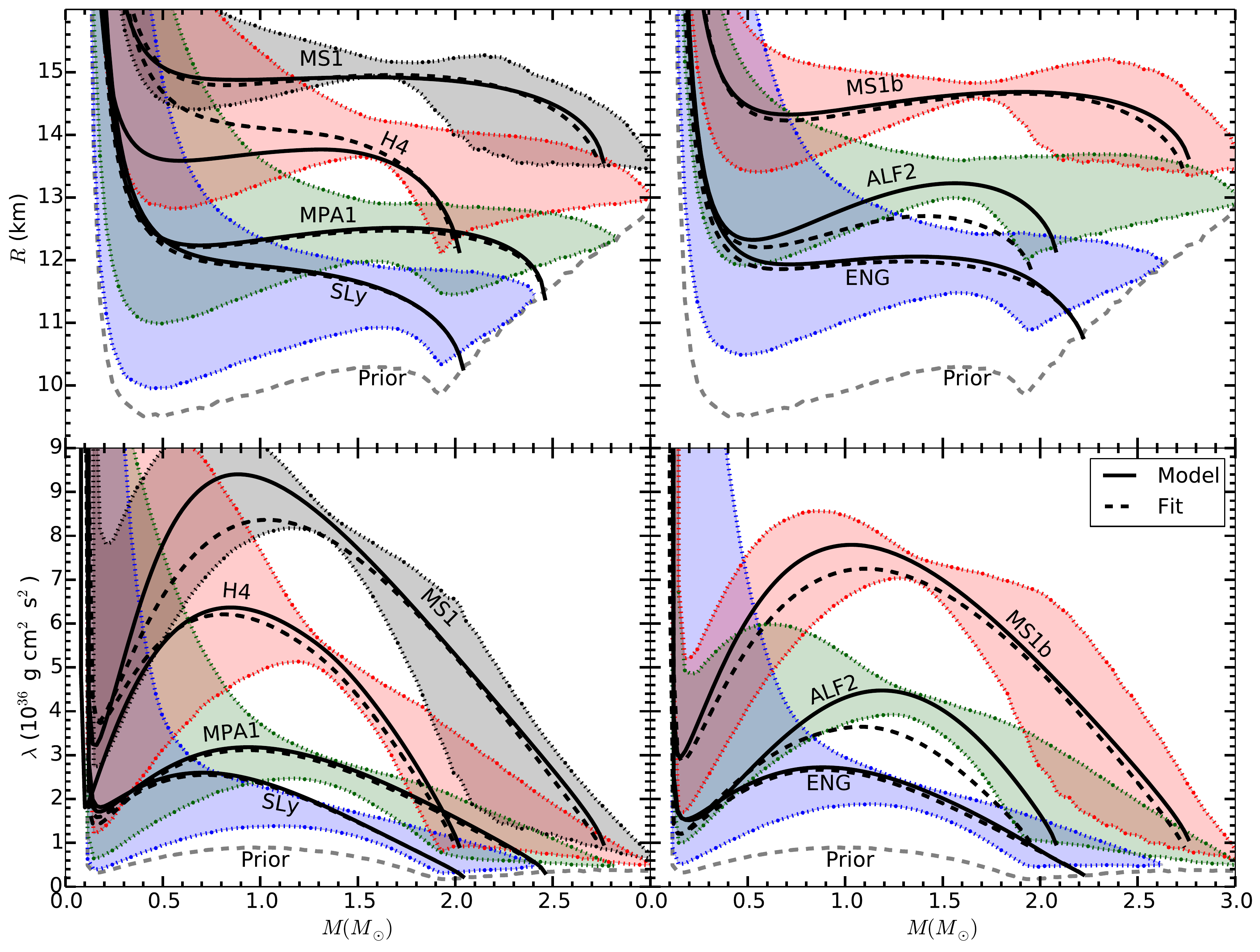}
\caption{95\% credible regions for the radius and tidal deformability for the same BNS systems and tabulated EOSs used in Fig.~\ref{fig:perroralleos}. In the left panels the EOSs are SLy, MPA1, H4, MS1 from bottom to top, and in the right panels they are ENG, ALF2 and MS1b from bottom to top. The tabulated EOS models, and not the fits, were used in generating the simulated waveforms. The lower bound from the mass and causality prior is also shown.}
\label{fig:structurealleos}
\end{center}
\end{figure*}

The corresponding uncertainties in $R$ and $\lambda$ for these 7 EOSs is shown in Fig.~\ref{fig:structurealleos}. The full width of the 95\% credible regions is $\sim 1$--2~km for the radius and $\sim 1$--$2\times 10^{36}$~g~cm$^2$~s$^2$ for the tidal deformability in the mass range $1M_\odot$--$2M_\odot$, and is roughly consistent for all EOSs. For $\lambda$, the credible region is smallest in the range $1.2M_\odot$--1.6$M_\odot$ from which the BNS population is sampled.

\subsection{Noise realizations and sampled populations}

\begin{figure}[!htb]
\begin{center}
\includegraphics[width=3.2in]{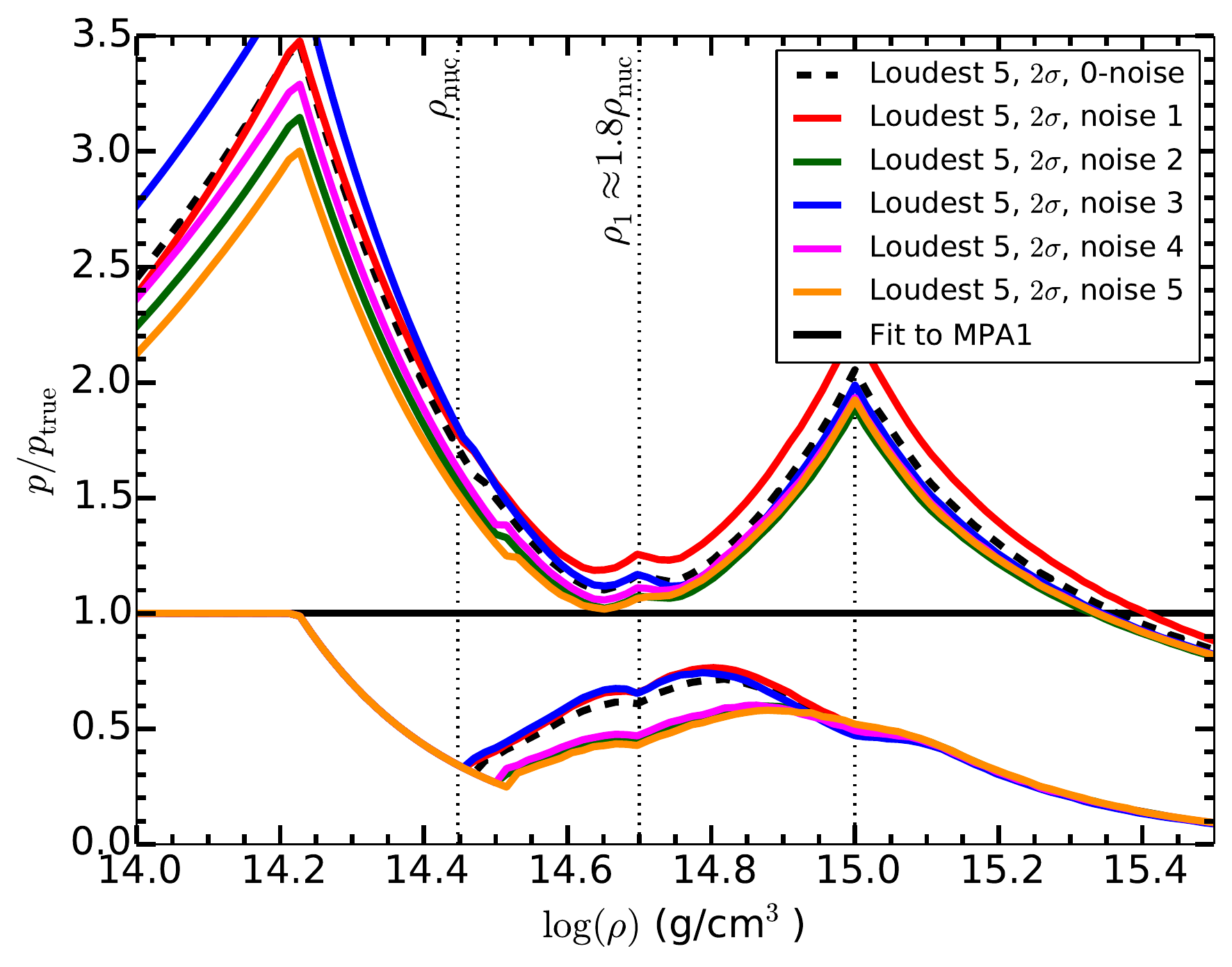}\\
\includegraphics[width=3.2in]{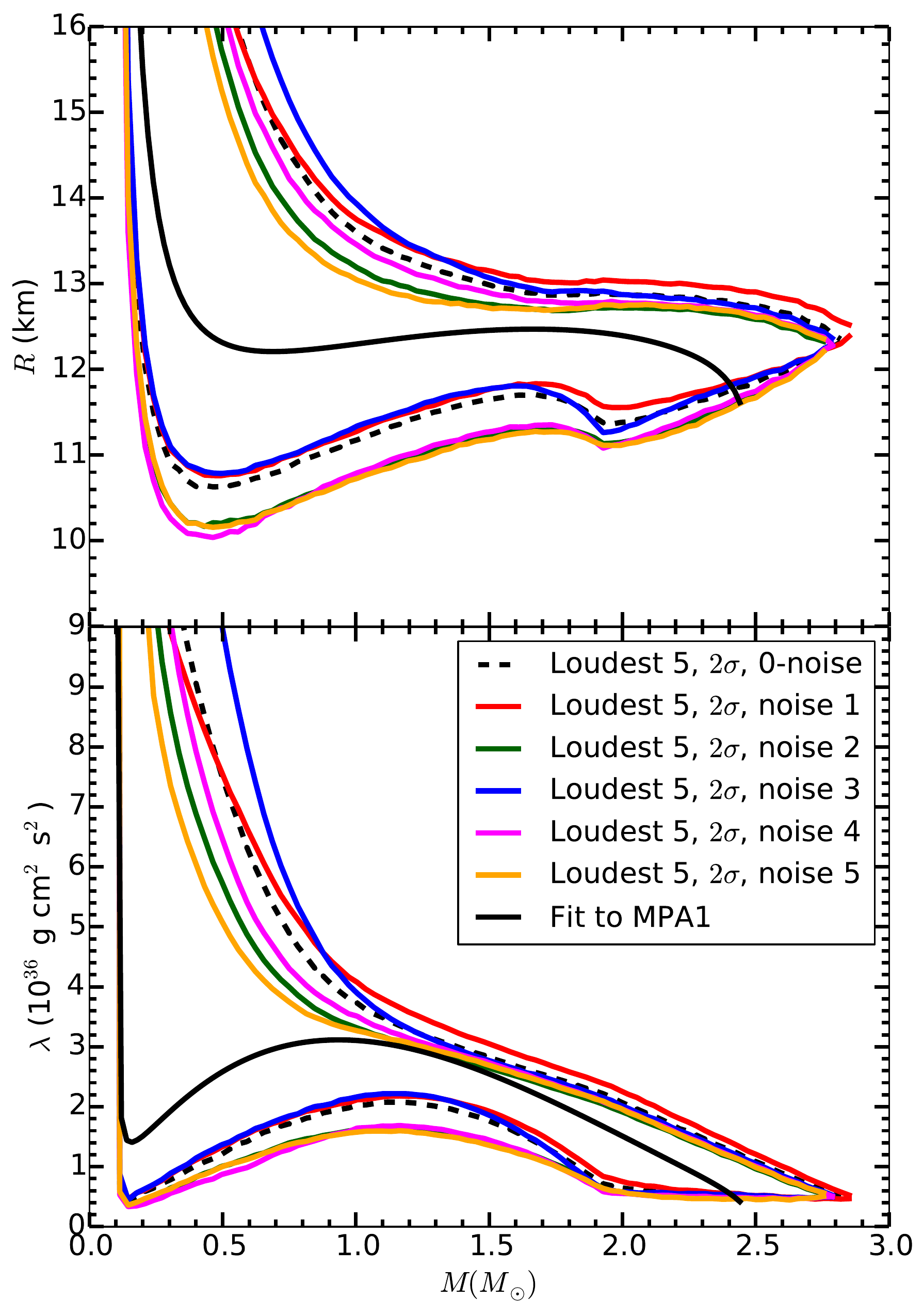}\caption{Same BNS population and priors as Figs.~\ref{fig:mpa1fit} and~\ref{fig:mpa1fitofn}. Contours represent 95\% credible regions for the loudest 5 events. The different contours correspond to different noise realizations. The dashed contour corresponds to the zero-noise data also shown in Fig.~\ref{fig:mpa1fitofn}.}
\label{fig:noise}
\end{center}
\end{figure}

In the above results, we used the zero-noise averaged likelihood (Eq.~\eqref{eq:averagedlike}) which gives results averaged over individual realizations of the detector noise. To determine how much the recovered EOS, radius, and tidal deformability can vary with the individual noise realizations, we injected the inspiral waveforms in our population into five different sets of detector noise and recovered the parameters with \texttt{LALInferenceMCMC}. Fig.~\ref{fig:noise} shows the recovered EOS, radius, and tidal deformability for each of our five noise realizations. For the pressure, the lower limits are roughly the same except in the region $10^{14.5}$--$10^{14.9}$~g/cm$^3$. This results from the fact that much of the lower bound is determined by the prior and the choice of EOS parameterization. We also show results for the zero-noise data presented in Fig.~\ref{fig:mpa1fitofn}, and this does appear to be the average of the individual noise realizations.

\begin{figure}[!htb]
\begin{center}
\includegraphics[width=3.2in]{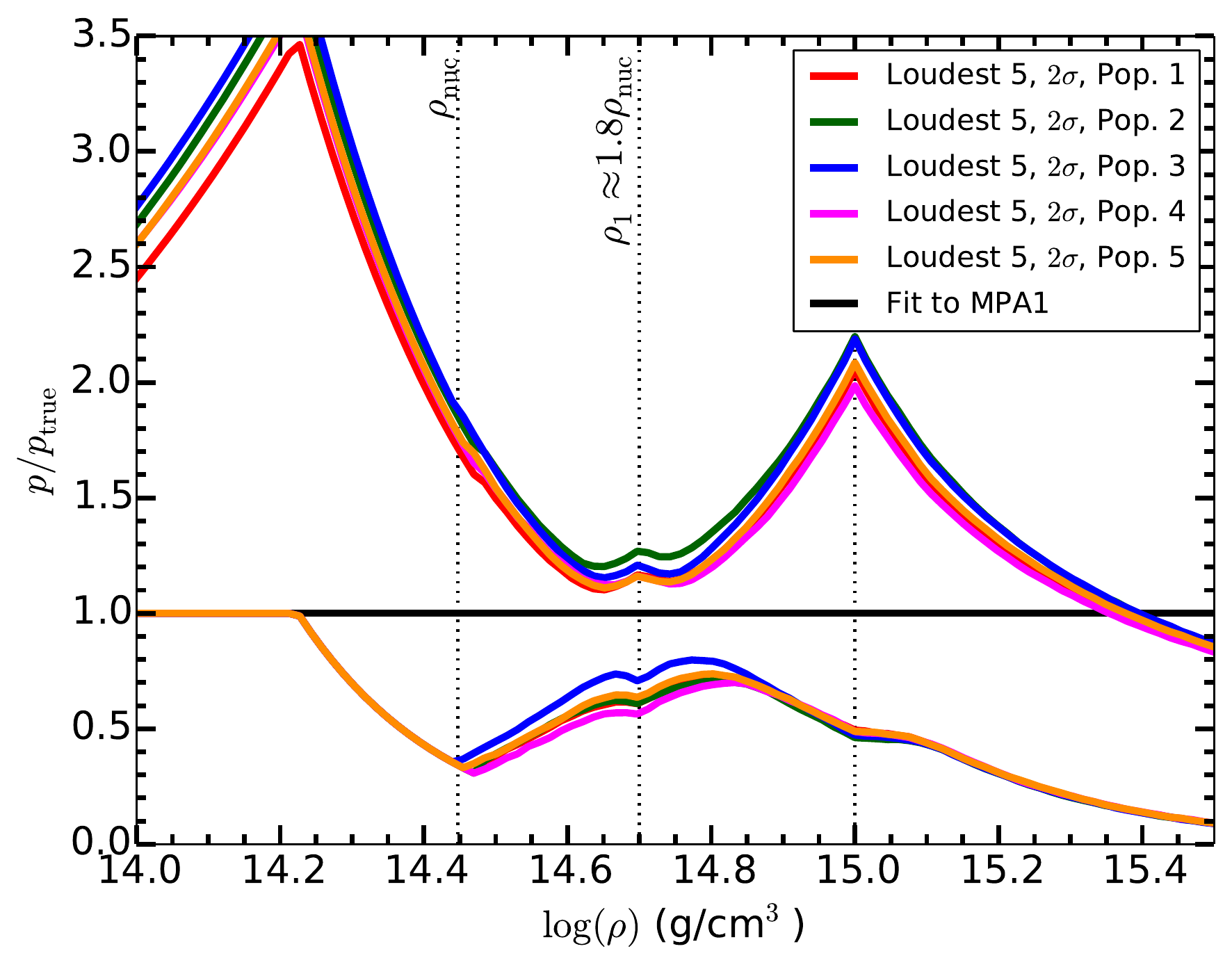}\\
\includegraphics[width=3.2in]{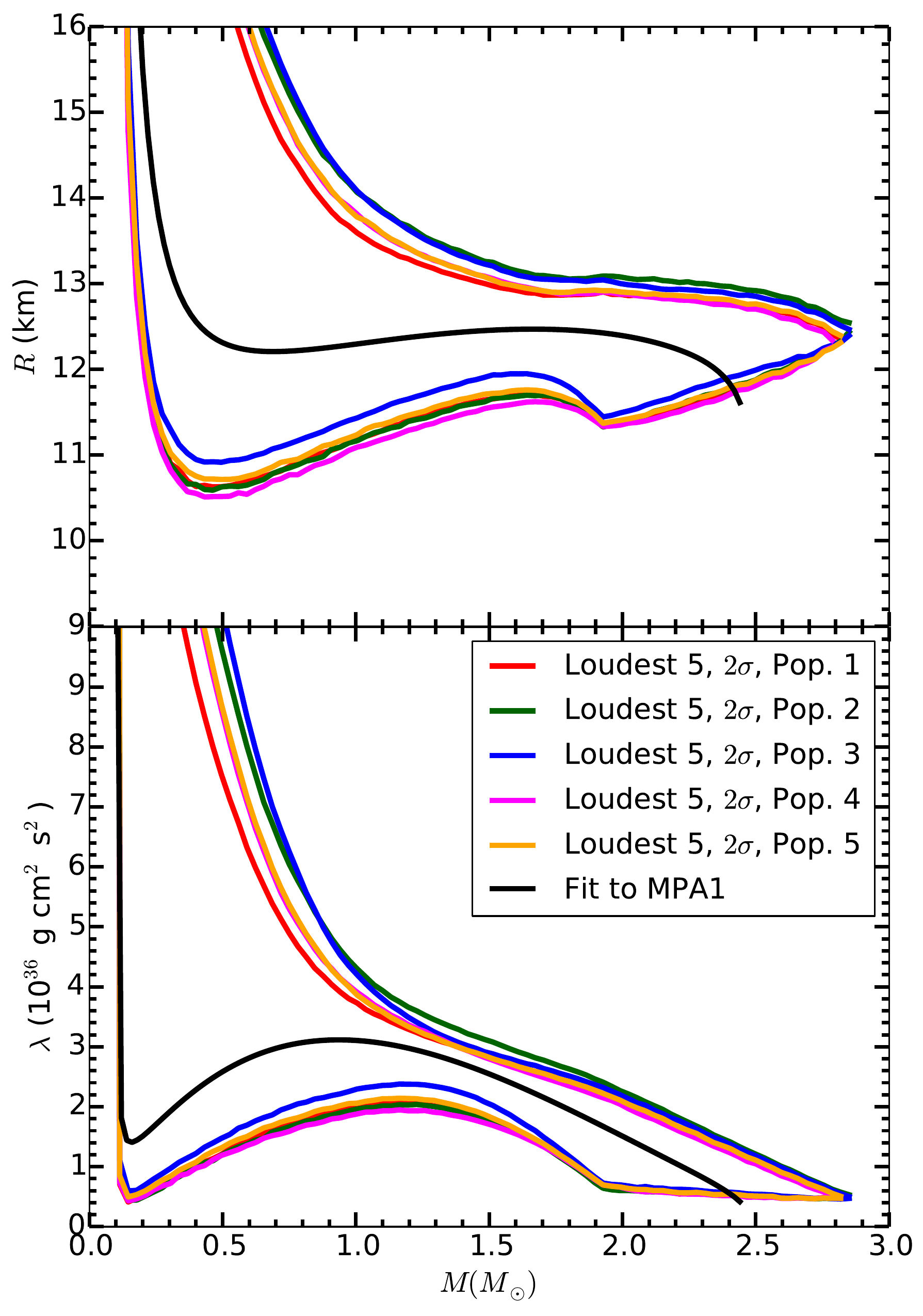}\caption{Five different populations of simulated BNS events for a one year period using the ``realistic'' event rate. Contours represent the 95\% credible regions for the loudest 5 events in each population using zero-noise data. Population 1 is the same as in Figs.~\ref{fig:mpa1fit} and~\ref{fig:mpa1fitofn}.}
\label{fig:populations}
\end{center}
\end{figure}

Finally, we examined how much the results depend on the sampled population. We sampled five different populations from the distribution described in Section~\ref{sec:baseline}. These populations had between 121 and 127 events with network SNR $\ge 8$, and the highest network SNR event in each population was between 41 and 88. We then evaluated the credible regions for the EOS, radius, and tidal deformability using zero-noise data. The results for the loudest five events in each population are shown in Fig.~\ref{fig:populations}. For a year of data with the ``realistic'' event rate, the reconstructed EOS as well as radius and tidal deformability are only mildly sensitive to the particular realization of the number of BNS events and source parameters.

\section{Systematic errors}
\label{sec:systematic}

\subsection{Waveform model}

The presentation of statistical errors above assumed an exact waveform model and a parameterized EOS that exactly fits the true EOS. At present, however, the PN inspiral waveform is only known completely to 3.5PN order in the point-particle terms, while the leading EOS-dependent tidal term enters at the same order as 5PN point-particle terms. Failing to include 4PN and higher-order point-particle terms can therefore bias the recovered parameters. As demonstrated in Fisher matrix studies~\cite{Favata2014, YagiYunes2014} and in an MCMC study~\cite{WadeCreightonOchsner2014}, the systematic error in the tidal parameter $\tilde\Lambda$ from the current waveform uncertainty is as large as the statistical error for aLIGO. 

\begin{figure}[!htb]
\begin{center}
\includegraphics[width=3.2in]{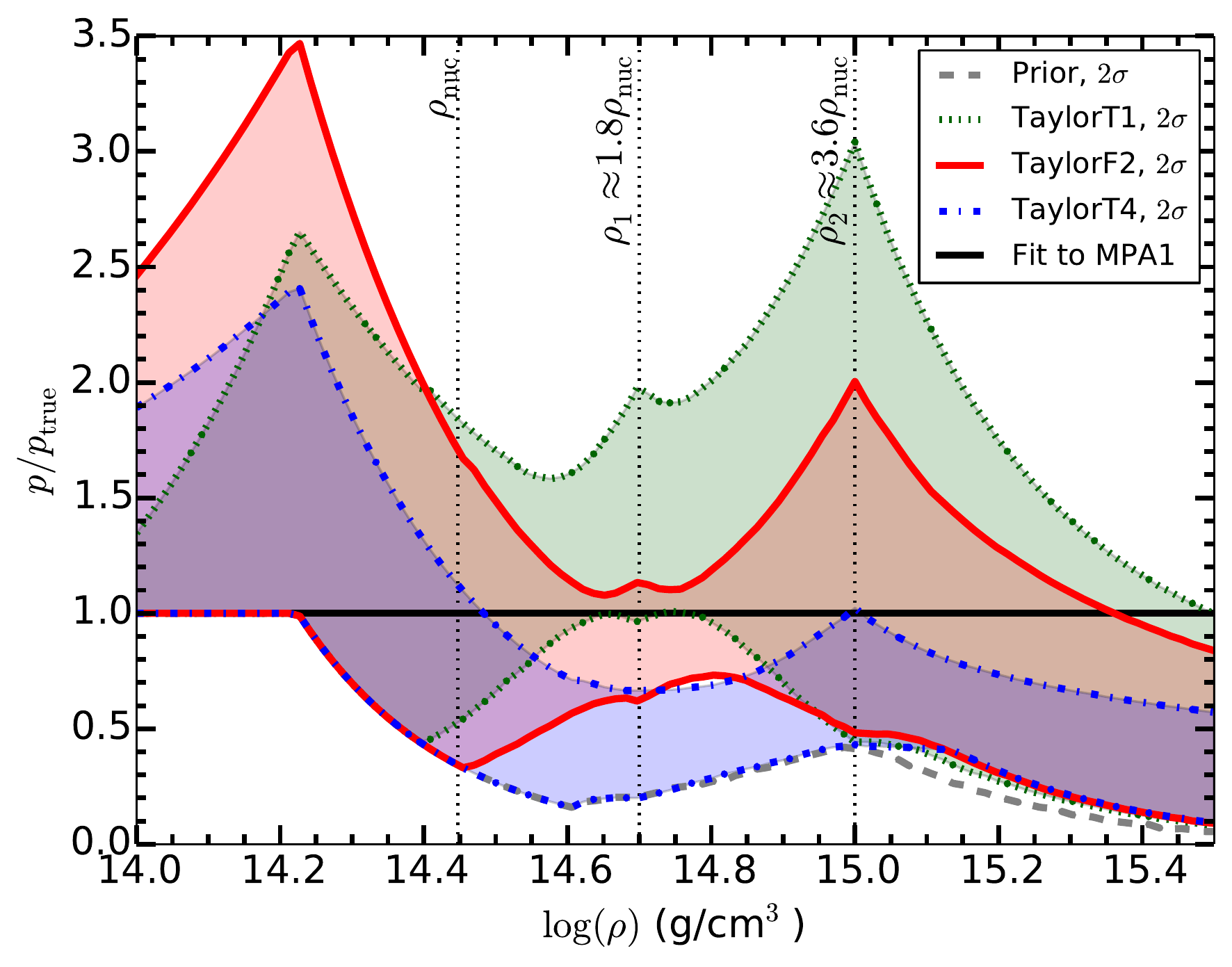}\\
\includegraphics[width=3.2in]{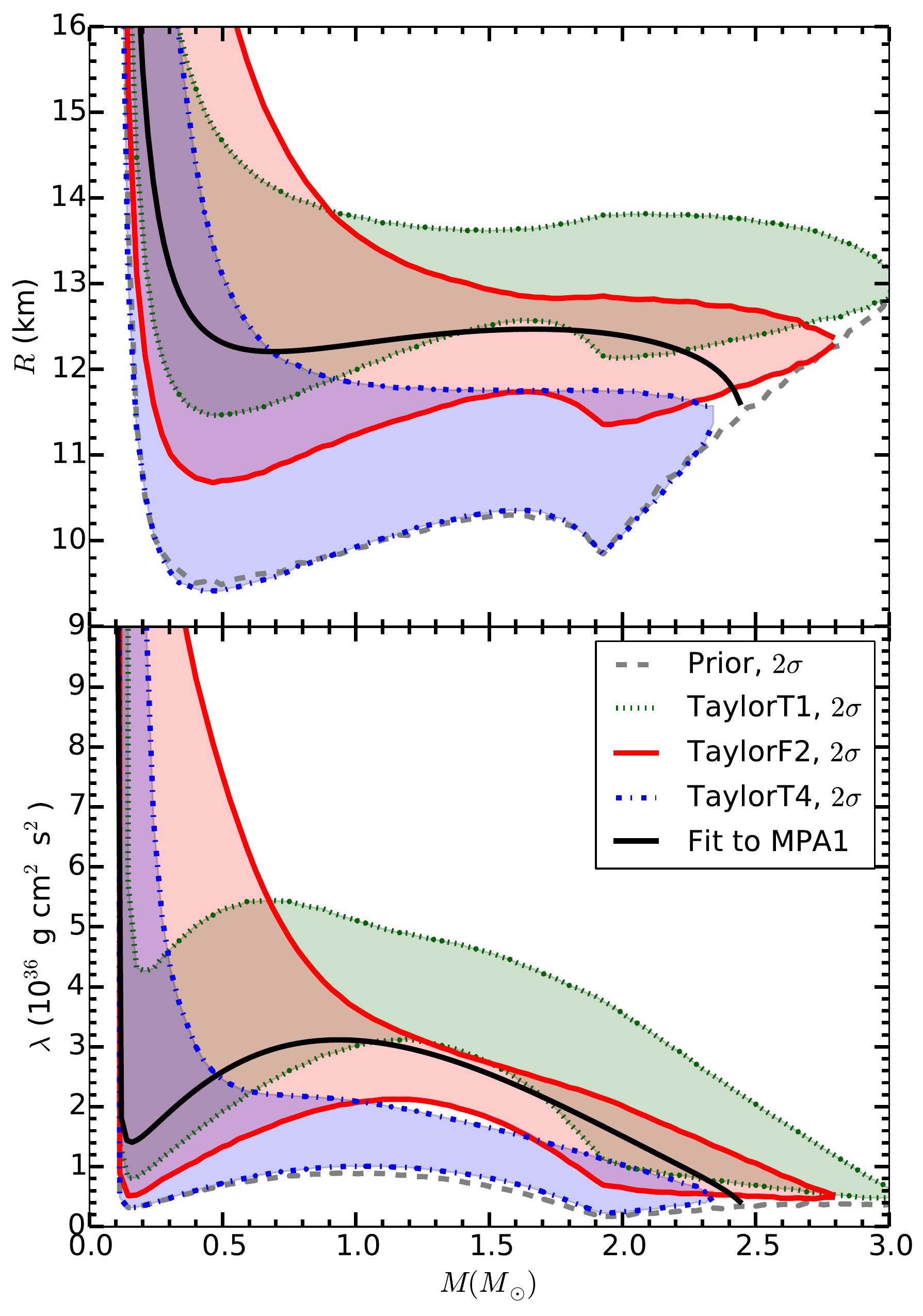}
\caption{Systematic errors in the recovered EOS due to uncertainty in the correct GW model. TaylorF2, TaylorT1, and TaylorT4 waveforms were injected into the data, and the parameters were recovered with the TaylorF2 waveform. The 95\% credible regions are shown for the loudest 5 systems.}
\label{fig:systematic}
\end{center}
\end{figure}

To examine the effect that uncertainty in the waveform model has on the recovered EOS, we will use a similar analysis to these previous works. We use variations in the current PN waveforms which vary only in how the waveform phase is calculated from the energy and luminosity. The waveform variations we use, described in Section~\ref{sec:waveform}, are the TaylorF2, TaylorT1, and TaylorT4 waveforms. For the loudest five events, we injected these three waveform variations into zero-noise data, then used the TaylorF2 waveform as the template to estimate the waveform parameters $\vec\theta$ with \texttt{LALInferenceMCMC}. In Fig.~\ref{fig:systematic} we show the bias in the recovered EOS, radius, and tidal parameter that results from the waveform uncertainty. The ordering is consistent with that of Ref.~\cite{WadeCreightonOchsner2014}; injecting the TaylorT1 waveform and recovering with the TaylorF2 waveform overestimates the tidal parameter while injecting the TaylorT4 waveform and recovering with the TaylorF2 waveform underestimates the tidal parameter. Likewise, injecting the TaylorT1 waveform overestimates the true pressure and radius, while the TaylorT4 waveform underestimates the true pressure and radius. Overall, failing to include the correct 4PN and higher point-particle terms can lead to a bias that is in some cases larger than the 95\% statistical uncertainty.

As in Refs.~\cite{Favata2014, YagiYunes2014, WadeCreightonOchsner2014}, we focused on the uncertainties in the point-particle description. However, other matter effects in addition to the quadrupole tidal interaction used here may also need to be accounted for. Ref.~\cite{Yagi2014}, for example, calculated the correction to the PN waveform from higher multipole tidal interactions. In addition, the amplification of the tidal deformation that occurs due to resonance when the GW frequency approaches each NS's f-mode frequency also leads to a small correction~\cite{FlanaganHinderer2008}. These effects are small, but will lead to a fractional error in the recovered parameters if not properly included. 

As an alternative to the PN approximation, the effective one body (EOB) formalism, which uses various techniques to re-sum the PN series, may converge more rapidly to the true binary waveform. EOB waveforms have been shown to accurately reproduce numerical binary black hole (BBH) waveforms. For example, Damour {\it et al.}~\cite{DamourNagarBernuzzi2013} compared a recent EOB implementation with nonspinning BBH simulations of the last $\sim 30$ GW cycles and found a phase difference of $<0.1$~radians after fitting the unknown 5PN contribution in the EOB radial potential to the numerical data. Since the tidal contribution to the waveform over this same interval is usually more than a radian, the current EOB waveform may be accurate enough for removing systematic errors due to uncertainties in the point-particle model. Tidal interactions have also been calculated in the EOB formalism for the first few multipoles to 2PN order in the EOB radial potential~\cite{BiniDamourFaye2012}, and comparisons with numerical BNS simulations have shown that EOB waveforms are consistent with the numerical waveforms, but only after calibration of currently unknown terms~\cite{BernuzziNagarThierfelder2012, HotokazakaKyutokuShibata2013}. Unfortunately, BNS codes are not currently as accurate as BBH codes, so waveforms calibrated with numerical BNS waveforms may still bias the recovered tidal deformability.

In our analysis we injected waveforms with zero NS spin and zero eccentricity and used a non-spinning, non-eccentric waveform template to recover the parameters. While NSs in known BNS systems have dimensionless spins of $|\chi| \lesssim 0.02$, not including the spin terms in the template can lead to systematic errors in $\tilde\Lambda$ that are greater than the statistical errors if the NSs have spins of $|\chi| \gtrsim 0.03$~\cite{Favata2014}. Likewise, the systematic error in $\tilde\Lambda$ from not including the eccentricity terms will be greater than the statistical error if the BNS system has an initial eccentricity at 10~Hz of $e_0 \gtrsim 0.003$, and this may occur for a small fraction of BNS systems formed in dense stellar environments~\cite{Favata2014}. These terms will need to be included in future studies.

Finally, in order to measure the NS EOS with BNS inspiral observations, one needs to know that one is observing a BNS inspiral. Attempting to recover the tidal parameter with a BNS waveform template, when one has actually observed a BHNS or BBH inspiral, will give erroneous results. One can positively identify a BH if its mass or spin is large enough because NSs have a more restricted range of allowed masses and spins ($M \lesssim 3.2M_\odot$ to satisfy causality~\cite{RhoadesRuffini1974} and dimensionless spin $| \chi | \lesssim 0.7$ to satisfy the Kepler constraint~\cite{CookShapiroTeukolsky1994}). Furthermore, NSs in known BNS systems have masses in the range $1.0M_\odot \lesssim M \lesssim 1.6M_\odot$ shown in Fig.~\ref{fig:rlofm} and spins $|\chi| \lesssim 0.02$~\cite{Favata2014}.  Restricting their analysis to aligned-spin systems, Hannam {\it et al.}~\cite{HannamBrownFairhurst2013} found that it would be difficult to distinguish between NSs and BHs in binaries except for the loudest few percent of signals. However, precession can break the mass-ratio--spin degeneracy, and using precessing waveforms, Chatziioannou {\it et al.}~\cite{ChatziioannouCornishKlein2014} found that one can distinguish between NSs and BHs for the majority of detected signals. Fortunately, as shown in Section~\ref{sec:baseline}, the majority of EOS information comes from the loudest few events, so even if one cannot determine if the weakest signals come from a BBH, BHNS, or BNS event, this will not significantly effect one's ability to measure the EOS.

\subsection{EOS fit}
\label{sec:systematicEOS}

In addition to errors in the waveform model, the choice of EOS parameterization can also effect the measurement of the EOS, and if unable to sufficiently capture complex behavior in the true EOS, the parameterization will introduce systematic errors. As shown in Fig.~\ref{fig:perroralleos}, the piecewise-polytrope fit can usually reproduce the tabulated EOS to a few percent. However, at the lower and upper density regions, the fit becomes worse. In addition, for the ALF2 EOS, there is a significant change in the EOS around $10^{14.5}$~g/cm$^3$ which is not modeled well by the fixed polytrope in that density interval. As a result, although the statistical error is small, the offset from the tabulated ALF2 EOS around $10^{14.5}$~g/cm$^3$ is larger than the 95\% credible interval for the statistical error. For EOSs such as ALF2, the 4-parameter piecewise polytrope used here will not be an appropriate fit. As found in Table~\ref{tab:eosfit}, it has the largest residual of the 7 EOSs by more than a factor of 2, mainly due to the behavior around $10^{14.5}$~g/cm$^3$. 

The recovered $R(M)$ and $\lambda(M)$ curves, however, are less effected by the inability of the EOS fit to capture detailed behavior of the tabulated EOSs. As can be seen in Fig.~\ref{fig:structurealleos}, when the tabulated EOS model is used as the ``true'' EOS, the tabulated EOS is still contained in the 95\% credible region. This is likely because, although the EOS fit poorly reproduces the tabulated EOSs at specific densities, integrating the stelar structure equations to find $R$ and $\lambda$ effectively smooths over these poorly fit density regions.

The original intent of the four-parameter piecewise polytrope was to provide a reasonable fit to a wide range of EOS models with a small number of parameters that have an intuitive meaning~\cite{ReadLackey2009}. However, it appears that for directly measuring the EOS with inspiral observations, a more sophisticated EOS will be needed. Two possibilities, mentioned in Section~\ref{sec:eosparam}, are the parameterization in Ref.~\cite{SteinerLattimerBrown2010}, which uses a piecewise polytrope with variable density intervals, and the spectral fit of Ref.~\cite{Lindblom2010}. However, the optimal parameterization for recovering the EOS from GW data remains to be determined.

\subsection{Other potential sources of error}

In addition to the systematic errors from the waveform model and EOS fit which can be reduced with better modeling, other effects that could interfere with the measurement of $\tilde\Lambda$ and the EOS have been suggested. Using the unipolar inductor model of Goldreich and Lynden-Bell~\cite{GoldreichLyndenBell1969}, it was suggested that interactions between a NS with a large magnetic dipole and its NS companion could create a torque on the binary that could potentially lead to a phase shift in the waveform of a few cycles~\cite{Piro2012}. If true, this would likely contaminate the measurement of tidal interactions. However, as pointed out by Lai~\cite{Lai2012}, the current in the circuit between the binary pair will generate a toroidal magnetic field that breaks the circuit leading to a maximum current. Lai then found that this upper limit is orders of magnitude too small to effect the waveform of the binary.

The tidal interaction discussed so far in this paper assumes the tidal field changes adiabatically. However, if the frequency of the tidal field variation (a multiple of the orbital frequency) approaches the resonant frequency of the various NS oscillation modes, the mode can be driven to large amplitudes. If the mode couples strongly to the tidal potential, this can lead to a phase shift as the binary passes through that resonance during the inspiral. This effect was examined for the g-modes, f-modes, and r-modes of nonspinning and spinning NSs~\cite{Lai1994, HoLai1999}. In general, the g-modes and r-modes will lead to a phase shift less than 0.1~radians, with maximum values for the largest spins and radii. The f-mode that has a small impact on the last few cycles, as discussed above, can lead to an appreciable effect at lower GW frequencies only if the NS is spinning at several hundred Hz. 

Finally, detector calibration errors will also impact the recovery of parameters. The uncertainty in the amplitude and phase of the Fourier transformed detector output $\tilde d(f)$ due to calibration errors are frequency dependent, and, for initial LIGO, were $\sim 10\%$ for the amplitude and $\sim 0.05$~radians for the phase of $\tilde d(f)$ over the bandwidth of the detector~\cite{VitaleDelPozzoLi2012}. The intrinsic parameters we are interested in here (masses and tidal parameters) are mostly determined by the phase evolution, and under the assumption that calibration errors will be similar for aLIGO, Vitale {\it et al.} found, using a Bayesian analysis, that the uncertainties in the chirp mass and symmetric mass ratio due to calibration errors would be $\sim 20\%$ as large as the statistical errors from detector noise~\cite{VitaleDelPozzoLi2012}. For the tidal parameter $\tilde\Lambda$ which contributes a few radians to the GW phase, this $\sim 0.05$~radian calibration error is not likely to dominate over the statistical error, but it should be included in future studies.

\section{Discussion}
\label{sec:discussion}

We have shown that, with a realistic population of BNS inspiral events, the advanced LIGO--Virgo GW network now undergoing construction can provide valuable information about the EOS and NS properties such as the radius and tidal deformability. Typical statistical errors in the pressure will be on the order of 10\% to a factor of 2 from 1--4 times nuclear density. This corresponds to a 95\% credible region of width $\sim 1$--2~km for the radius and $\sim 1$--$2\times 10^{36}$~g~cm$^2$~s$^2$ for the tidal deformability over the mass range $1M_\odot$--$2M_\odot$. These results are in agreement with those of Ref.~\cite{DelPozzoLiAgathos2013} which showed that with tens of BNS observations, $\lambda$ could be measured at a reference mass of $1.4M_\odot$ to about 10\%. However, our results show that incorporating additional known EOS information allows us to measure NS properties at other masses as well, and this is true even if BNS systems have a narrow range of NS masses.

However, in Section~\ref{sec:systematic} we found that currently unknown high order point-particle PN terms will create a large systematic bias in the recovered EOS that will be larger than the statistical uncertainties. These terms will need to be calculated or fit with numerical simulations to accurately measure the EOS, and significant work remains before the resulting waveform templates can be trusted. The choice of EOS parameterization can also effect the recovered EOS, and EOSs with complex behavior such as the ALF2 EOS will need to be modeled with more sophisticated EOS parameterizations. 

Our results indicate that advanced GW detectors can measure the NS radius with similar statistical errors to those from electromagnetic observations of NSs whose errors are typically in the range of a few km. However, although both electromagnetic and GW measurements have systematic errors, the gravitational waveforms from inspiralling BNS systems are an intrinsically cleaner source of data than the electromagnetic emission of NSs. Ultimately, electromagnetic and GW measurements will have to agree, and comparing the results will provide an important consistency check on our understanding of general relativity, electromagnetic radiation from NSs, and the EOS.
 
\acknowledgements

We sincerely thank Walter Del Pozzo, Tyson Littenberg, and Jolien Creighton for advice and corrections, as well as Evan Ochsner for help in generating the parameters of the BNS populations used here. BL also benefited from discussions with Will Farr, Tjonnie Li, and Chris Van Den Broeck and from discussions at the University of Washington's Institute for Nuclear Theory Program 14-2a. This work was supported by NSF Grants PHY-1305682, PHY-0970074, PHY-0955929, and AST-1333142. Simulations were run on the Nemo cluster at the University of Wisconsin--Milwaukee, supported by the NSF Grant PHY-0923409, and the Orbital cluster at Princeton University.

\bibliography{paper}
\end{document}